\documentclass[12pt]{article}
\usepackage[dvips]{graphicx}
\usepackage{epsfig}
\usepackage{amsfonts}
\usepackage{amsmath}
\usepackage{graphics,color}
\begin{document}
\def\ren{renomaliza}
\def\k{\kappa}
\def\ti{\tilde}
\def\da{\dagger}
\def\a{\alpha}
\def\b{\beta}
\def\g{\gamma}
\def\l{\lambda}
\def\d{\delta}
\def\D{\Delta}
\def\p{\partial}
\def\t{\theta}
\def\s{\sigma}
\def\G{\Gamma}
\let\vv\v
\def\v{\varepsilon}
\def\m{\mu}
\def\n{\nu}
\def\tr{\textrm}
\def\tir{\tilde{r}}
\def\nn{\nonumber}
\def\w{\wedge}
\def\hw{\hat{\omega}}
\def\K{K\"ahler}
\def\o{\omega}
\def\calu{{\cal U}}
\def\calf{{\cal F}}
\def\Tr{{\rm Tr}}
%%%%%%%%%%%%%%%%%%%%%%%%%%%%%%%%%%%%%%%
%%%%%%%%%%%  Manavendra %%%%%%%%%%%%%%%%%%%%%%%
\def\ti{\tilde}
\def\da{\dagger}
\def\a{\alpha}
\def\b{\beta}
\def\g{\gamma}
\def\l{\lambda}
\def\d{\delta}
\def\k{\kappa}
\def\p{\partial}
\def\t{\theta}
\def\s{\sigma}
\def\G{\Gamma}
\def\v{\varepsilon}
\def\tr{\textrm}
\def\nn{\nonumber}
\def\w{\wedge}
\def\hw{\hat{\omega}}
\def\K{K\H{a}hler}
\def\cf{{\cal F}}
\def\o{\omega}
%\begin{document}
\newcommand{\be}{\begin{equation}}
\newcommand{\ee}{\end{equation}}
\newcommand{\bea}{\begin{eqnarray}}
\newcommand{\eea}{\end{eqnarray}}
\pagenumbering{arabic}
%\renewcommand{\theequation}{\thesection.\arabic{equation}}

%%%%%%%%%%%%%%%%%%%%%%%%%%%%%%%%%%%%%%%
%%%%%%%%%%%  Sergio %%%%%%%%%%%%%%%%%%%%%%%
\def\ren{renomaliza}
\def\ti{\tilde}
\def\da{\dagger}
\def\a{\alpha}
\def\b{\beta}
\def\g{\gamma}
\def\l{\lambda}
\def\d{\delta}
\def\p{\phi}
\def\t{\theta}
\def\s{\sigma}
\def\m{\mu}
\def\n{\nu}
\def\tr{\textrm}
\def\tir{\tilde{r}}
\def\nn{\nonumber}
\def\w{\wedge}
\def\hw{\hat{\omega}}
\def\K{K\"ahler}
\def\o{\omega}
\def\calu{{\cal U}}
\def\Tr{{\rm Tr}}
\newcommand{\IC}{\mathbb{C}}
\newcommand{\IP}{\mathbb{P}}
\newcommand{\IZ}{\mathbb{Z}}
\newcommand{\IR}{\mathbb{R}}
\newcommand{\IF}{\mathbb{F}}
\newcommand{\ad}{\dot{\a}}
\newcommand{\bd}{\dot{\b}}

\numberwithin{equation}{section}

\begin{titlepage}

%\version\versionno

\hfill hep-th/0512061

\hfill MCTP-05-97

\hfill  UT-05-18

\hfill NSF-KITP-05-105

\begin{center}
%\vskip 2.5 cm
\vskip 1 cm
{\Large \bf The Gauge/Gravity Theory of Blown up Four Cycles}
\vskip .3cm
%{\Large \bf }

 \vskip .7 cm
 \centerline{ {\large Sergio Benvenuti$^{1}$, Manavendra Mahato$^{2}$,}}

 \vskip .5 cm

 \centerline{ {\large  Leopoldo A. Pando Zayas$^{2}$  and Yuji Tachikawa$^{3,4}$}}

  \vskip 1 cm

\end{center}
\vskip .4cm \centerline{\it ${}^1$ Scuola Normale Superiore, Pisa}
\centerline{ \it and INFN, Sezione di Pisa, Italy }

 \vskip .4cm
\centerline{\it ${}^2$ Michigan Center for Theoretical Physics}
\centerline{ \it Randall Laboratory of Physics, The University of
Michigan} \centerline{\it Ann Arbor, MI 48109-1120}

\vskip .4 cm
 \centerline{\it ${}^3$ Department of Physics, Faculty of Science,}
 \centerline{ \it University of Tokyo, Tokyo 113-0033, JAPAN}

\vskip .4 cm
 \centerline{\it ${}^4$ Kavli Institute for Theoretical Physics,}
 \centerline{ \it University of California, Santa Barbara, CA 93106, USA}

\vskip 1.5 cm
\begin{abstract}

We present an explicit supersymmetric deformation of supergravity
backgrounds describing D3-branes on Calabi-Yau cones. From the
geometrical point of view, it corresponds to blowing up a 4-cycle
in the Calabi-Yau and  can be done universally. In the field theory, we identify this
deformation with motion on non-mesonic directions in the full
moduli space of vacua. For the case of a $\mathbb{Z}_2$ orbifold
of the conifold, we discuss an explicit gravity solution with two
deformation parameters: one corresponding to blowing up a
$2$-cycle and one corresponding to blowing up a $4$-cycle. The
generic case where the Calabi-Yau is toric is also discussed in
detail. Quite generally, the order parameter of these $4$-cycle
deformations is a dimension six operator. We also consider probe
strings which show linear confinement and probe D7 branes which
help in understanding the behavior far in the infrared.
\end{abstract}

\end{titlepage}

%%%%%%%%%%%%%%%%%%%%%%%%%%%%%%%%%%%%%%%%%%%%%%%%%%%%%%%%%%%%%%%%%%%%%%%%%%%%%%
\section{Introduction}
%%%%%%%%%%%%%%%%%%%%%%%%%%%%%%%%%%%%%%%%%%%%%%%%%%%%%%%%%%%%%%%%%%%%%%%%%%%%%%%
The AdS/CFT correspondence has provided an alternative approach to
the study of gauge theories at large 't Hooft coupling
\cite{malreview}. One of the most powerful properties is that the
correspondence provides a prescription for its deformation via the
state/operator correspondence \cite{wittengkp}. One interesting
line of development has focused on examples with minimal
supersymmetry in four dimensions. This corresponds to backgrounds
of the form $AdS_5 \times X^5$, where $X^5$ is Sasaki-Einstein.
The prototypical example of $X^5$ is $T^{1,1}$ \cite{kw}.
Recently, an infinite class of explicit Sasaki-Einstein metrics
called $Y^{p,q}$ \cite{se} has been found. The dual superconformal
quiver gauge theories have been identified in
\cite{Benvenuti:2004dy}. This class has then be generalized to the
$L^{p,q,r}$
 \cite{cvetic, lpqr,Benvenuti:2005ja,
Franco:2005sm, Butti:2005sw}.

In this paper, we discuss a deformation of the AdS/CFT
correspondence involving motion in the K\"ahler moduli space of
the Calabi-Yau cone over the Sasaki-Einstein base. Such effects
for blown up $2$-cycles have been studied for the resolved
conifold by Klebanov and Witten \cite{kw1}. The K\"ahler modulus
corresponds to giving vacuum expectation values (vev) to the
fundamental fields in such a way that no mesonic operator gets a
vev, but a real dimension $2$ scalar operator acquires a vev. We
can call this ``motion in the moduli space of vacua along
non-mesonic (or baryonic) directions''.

We will be interested in blowing up $4$-cycles. Global aspects of
the geometry restricts the list of spaces but one still finds a
large class including: $S^5/\mathbb{Z}_3, T^{1,1}/\mathbb{Z}_2$,
and appropriate orbifolds of some $Y^{p,q}$ and $L^{p,q,r}$. The
key observation is that locally there is a simple Calabi-Yau
deformation for every Calabi-Yau cone, corresponding to blowing up
a 4-cycle. Global issues restrict the examples somewhat, but there
we still have an infinite number of explicit examples.
Interestingly, this statement has a very rich history. Calabi
considered properties of holomorphic fibers over K\"ahler-Einstein
spaces in \cite{calabi}, and also established various properties
of such holomorphic fiber bundles including K\"ahlerness and
reduced $SU(N)$ holonomy. In a more explicit setting Page and Pope
\cite{pp} considered a very similar problem of constructing
Einstein metrics in dimension $n+2$ starting with a
K\"ahler-Einstein metric in dimension $n$.
%The concrete set of
%metrics we use in this paper then follows as a particular case of
%the analysis of Page and Pope.
This deformation has appeared in the context of the AdS/CFT in
some concrete examples though never recognized as universal. In
\cite{pt}, it was discussed as a concrete generalization for the
conifold and also its small resolution and complex deformation.
More recently, it has been discussed in the context of $Y^{p,q}$
\cite{pal} and $L^{p,q,r}$ \cite{sfetsos}. In this paper, we
emphasize its universal character and discuss various aspects of
the gauge theory duals.

The organization of the paper is as follows.

In section \ref{d3branes} we discuss the local geometry showing
that the blown-up four cycle is calibrated with the K\"ahler form
of the Calabi-Yau and it is, therefore, a divisor. In this section
we also consider placing D3 branes on the blown up $4$ cycle. We
write down the Calabi-Yau metric and the explicit near horizon
supergravity solution. Using a large radius expansion we argue
that in the dual field theory the blow up corresponds to giving a
vev to a dimension six operator.

We then discuss (section \ref{kaehler}) in detail a specific
example, the so called vanishing $\IP^1 \times \IP^1$ geometry.
The Calabi-Yau is a line bundle over the complex surface $\IP^1
\times \IP^1$. Interestingly, \cite{pt} found the metric for
arbitrary size of the \emph{two} $\IP^1$s.  When one $\IP^1$ has
zero size, this is a $\IZ_2$ orbifold of the well-understood
example of the small resolution of the conifold. In the
corresponding field theory we relate the two deformations to
motion in the moduli space of vacua. More precisely, we find two
non-mesonic directions in the moduli space of vacua corresponding
to the two K\"ahler parameters in the geometry. We also analyze a
double scaling limit of the metric of \cite{pt} which leads to the
Eguchi-Hanson metric. We finish section \ref{kaehler} with a
presentation of the toric description of the blow up in terms of
Gauged Linear Sigma Models. Although the treatment is specific to
a particular geometry, the techniques are universal and can be
extended to more complicated examples.

Section \ref{general} contains a careful discussion of various
global issues. In particular, we discuss the case of toric
Calabi-Yau's. A nice discussion of toric geometry can be found for
our case in \cite{MS}, where there is also a detailed analysis of
the $Y^{p,q}$ case. We also consider topological obstructions to
blowing up a 4-cycle. Using the $(p,q)$-web language, which is
dual to the toric language \cite{Aharony:1997bh, Leung:1997tw}, we
classify the K\"ahler deformations in ``local'' and ``global.'' This
terminology originated in \cite{Aharony:1997bh} are refers to the
energy required to perform the deformation, it suggests that the
order parameter has always dimension $2$ for the global
deformations and dimension $6$ for the local deformations.

A very natural way to perturb conformal quiver gauge theories
involve changing the ranks of some of the gauge groups. This line
started in \cite{kn} and lead to the Klebanov-Strassler solution
\cite{ks}. The supergravity dual of this generalization has been
constructed for spaces like $Y^{p,q}$ \cite{HEK, ben} and
$L^{p,q,r}$ \cite{lpqr, Gepner:2005zt}. In section \ref{cascade},
we show that blowing up a 4-cycle is compatible with introducing
fractional branes\footnote{Various examples appeared in the
literature treated as a case by case situation
\cite{pt,pal,sfetsos}.}, and explicitly see that the fractional
brane is now a wrapped D5 brane.

In a series of appendices we cover a number of technical issues
and explore the behavior of some probes in the geometry. In
appendix \ref{ricci} we work out all the metric information
including an explicit proof of Ricci flatness. A natural question
is about supersymmetry. This question can be answered in very
general grounds elaborating on arguments by Calabi \cite{calabi}
and Page and Pope \cite{pp}. We present an explicit calculation
whose details are offered in appendix \ref{susy}.
 Appendix \ref{action} presents all the details of a compactification given
in the main text that helps in the identification of the mass of
the supergravity mode involved in blowing up the four cycle. Some
global issues of our deformation applied to quasi-regular
Sasaki-Einstein spaces are treated in appendix \ref{orbifolding}.
We show that any quasiregular $Y^{p,q}$ admits the K\"ahler
deformation after a suitable orbifold. In appendix
\ref{confinement} we  use a classical probe string to see that the
supergravity background allows for a dual Wilson loop with area
law behavior, thus showing that the deformation induces
confinement.
%To understand some
%physical properties of the solution we consider two classical
%probes.
As a way to probe the singularity we consider a probe D7 brane in
appendix \ref{d7}.

%%%%%%%%%%%%%%%%%%%%%%%%%%%%%%%%%%%%%%%%%%%%%%%%%%%%%%%%%%%%%%%%%%%%%%%%%%%%%%
\section{Calabi-Yau metrics and D3 branes on blown up 4-cycles}\label{d3branes}
%%%%%%%%%%%%%%%%%%%%%%%%%%%%%%%%%%%%%%%%%%%%%%%%%%%%%%%%%%%%%%%%%%%%%%%%%%%%%%%
Consider a metric of the form
\be \label{lametrica} ds^2=
A^2(r)dr^2+B^2(r)r^2(d\psi+\s)^2+C^2(r)r^2 ds_4^2,
\ee
where the
four-dimensional space $M^4$ is K\"ahler-Einstein and $\s$ is such
that $d\s=2 J$, where $J$ is the  K\"ahler form on $M^4$. For a
simple case $A=B=C=1$, the metric is Ricci flat. This is also true
for the case when we deform the metric by a $b$
parameter\footnote{Page and Pope \cite{pp} studied the question of
when such metrics as (\ref{lametrica}) were Einstein.} as
\be
A^{-2}=B^2=1-\frac{b^6}{r^6}, \qquad C=1.
\ee The range of the
radial coordinate is now $b \leq r \leq \infty$. The quantity $b$ parameterizes
motion in K\"ahler moduli space of the Calabi-Yau. There are many
concrete examples in this class of metrics: $S^5$, $T^{1,1}$,
$Y^{p,q}$, $L^{p,q,r}$.

The four-cycle being blown up at $r=b$ is $ds_4^2$. It is
calibrated with respect to {\K} calibration $\frac{1}{2}J\wedge J$
at $r=b$. The {\K} form on the 6 dimensional manifold is
\be
J_6=r^2 J +rdr\wedge (d\psi +\s).
\ee
Its pullback on the 4-cycle is $J'=r_i^2J=r_i^2(e_1\w e_2+e_3\w
e_4)$ at a given $r=r_i$ in terms of veilbeins in orthonormal
basis. This leads to $\frac{1}{2}J'\w J' =r_i^4(e_1\w e_2\w e_3\w
e_4)$, which is same as the volume form of the four-cycle
calculated using the pullback metric at $r_i=b$.

Let us look more carefully at the geometry near $r=b$.  In order
to do so, we introduce a new radial coordinate given by
\be u^2=
\frac19 r^2 \left(1-\frac{b^6}{r^6}\right). \label{udef}
\ee
The metric then
becomes \be
ds^2= \frac{9\, du^2}{(1+2\frac{b^6}{r^6})^2}+ u^2
(d\psi+\sigma)^2 + r^2(u) ds_4^2. \ee
For $u\to 0 \,\,\, (r\to b)$,
we have approximately \be
ds^2\approx
du^2+9u^2 (d\psi+\sigma)^2 + b^2 ds_4^2. \ee
In order to have a complete metric, the periodicity of $3\psi$ has to be $2\pi$.
If this is the case, then (\ref{lametrica}) is an explicit Calabi-Yau
metric on resolved complex cones. We will discuss other global aspects of these
metrics in section \ref{general}.

%%%%%%%%%%%%%%%%%%%%%%%%%%%%%%%%%%%%%%%%%%%%%%%%
\subsection{D3 branes}
%%%%%%%%%%%%%%%%%%%%%%%%%%%%%%%%%%%%%%%%%%%%%%%%%%
To study the role of the above spaces in the context of the
AdS/CFT correspondence, we need to consider adding $N$ D3 branes
filling the four dimensional space-time then taking the decoupling
limit. Recall that given a six-dimensional Ricci-flat space we can
always construct a D3 brane solution of the form \bea \label{d3}
ds^2 &=& h^{-1/2} \eta_{\m\n}dx^\m dx^\n + h^{1/2} g_{mn}dy^m
dy^n,\nonumber \\
F_5&=& (1+\star)dx_0\wedge dx_1 \wedge dx_2 \wedge dx_3 \wedge dh^{-1}, \nonumber  \\
\label{met3} \Phi&=&\Phi_0. \eea The  Einstein's equations reduce
to: \be {}^6R_{mn}=0, \qquad {}^6 \bigtriangleup h=0. \ee
Consider first the conical solution, with $A=B=C=1$. Putting the D3 branes
at the apex of the cone and taking the near horizon limit, we
obtain spaces of the form $AdS_5\times X^5$, where $X^5$ is
Sasaki-Einstein, for instance one of the 5-d spaces enumerated
above. The field theory is explicitly known for this class of
backgrounds and it can be conveniently packed in a quiver diagram
plus a superpotential. In this paper we would like to consider the
effect on the field theory of turning on some small
$b$-deformation. In the case the D3 branes are smeared
homogeneously on the blown up $4$-cycle (all the D3 branes are at
$r=b$), the warp factor for the above class of solutions can be
written explicitly as \cite{pt}: \be \label{warp}
h=-\frac{2L^4}{b^4}\bigg[\frac16\ln
\left(\frac{(\tir^2-1)^3}{\tir^6-1}\right)
+\frac{1}{\sqrt{3}}\left(\frac{\pi}{2}-\arctan\left(\frac{2\tir^2+1}{\sqrt{3}}\right)\right)\bigg].
\ee where $\tir=r/b$. For large $r$ the warp factor becomes the
traditional $L^4/r^4$; the dependence on $b$ arises at the level
of $1/r^6$ corrections. In the limit of $r\to b$ the warp factors
is approximately $-(2L^2/3b^4)\ln(r/b -1)$.

%%%%%%%%%%%%%%%%%%%%%%%%%%%%%%%%%%%%%%%%%%%%%%%%%%%%%%%%%%%%%%%%%%%%%%%%%%%%%%
\subsection{Large radius expansion}
%%%%%%%%%%%%%%%%%%%%%%%%%%%%%%%%%%%%%%%%%%%%%%%%%%%%%%%%%%%%%%%%%%%%%%%%%%%%%%%

The process of blowing up a $4$-cycle implies that in the
superconformal quiver gauge theory scalar gauge invariant
operators are taking vevs.\footnote{There is also the logical
possibility that this corresponds to changing the parameters in
the Lagrangian. Relevant or exactly marginal deformations in
quiver theories are however highly constrained (see for instance
\cite{Benvenuti:2005wi}). Moreover they typically correspond to
turning on $3$-form field strength in the gravity. K\"ahler
deformations in the Calabi-Yau always correspond to scalar
operators taking vevs.} A general property of this process is that
the $R$-charge of the gauge invariant operators taking vev is
vanishing. The reason is that the backgrounds preserve an $U(1)$
symmetry associated  to the $R$-charge. This implies that no
chiral mesonic operator is taking a vev. This fact is expected,
since chiral mesonic field (motion in the mesonic branch)
parameterize the motion of D3 branes on the Calabi-Yau space, and
in our backgrounds the D3 branes are always at the minimum
possible value of $r$, i.e. $r=b$.

Among the infinite tower of operators turned on, the ones with
smallest conformal dimension are typically called order
parameters. For instance in the case of the small resolution of
the conifold (corresponding to blowing up a two cycle) the order
parameter is a supersymmetric partner of the baryonic current and
has dimension two, as explained in \cite{kw1}.\footnote{In Section
\ref{general} we will generalize this result to a generic K\"ahler
motion corresponding to blowing up two-cycles and four-cycles.}
We now show in detail, through a large $r$ expansion of the exact solution, that
for the $b$-deformation, the order parameters have dimension six.

An important entry in the AdS/CFT dictionary explains the
relationship between an operator ${\cal O}$ of conformal dimension
$\Delta$ in the CFT and solutions to the linearized gravity
equations. The solution, whose behavior at large values of $r$ is
\be \delta \phi =a r^{\Delta-4} + c r^{-\Delta}, \ee corresponds,
on the CFT side, to $H = H_{CFT} + a {\cal O}$, while $c = <{\cal
O}>$.

Let us consider the linearized form of the metric (\ref{met3}).
One can write it as \bea
ds^2&=&\frac{r^2}{L^2}dx_\mu dx^\mu + \frac{L^2}{r^2} dr^2 + L^2 ds^2(X_5) \nonumber \\
&-& \frac{b^6}{5L^2r^4} dx_\mu dx_\mu + \frac{L^2b^6}{5r^8}dr^2 +
\frac{L^2b^6}{5r^6} ds^2_4 -\frac{L^2b^6}{r^6}(d\psi +\sigma)^2+
\ldots \eea The first line is simply the metric on $AdS_5\times
X_5$. The second line gives the form of the linearized fields from
where we can read the transformation properties and the dimensions
of the corresponding field theory operators.  For the metric
perturbations, we find it convenient to normalize them by dividing
by the background value of the metric. This implies that the
modifications of the metric are due to a dimension six operator.

Let us check this result considering the other supergravity
fields. Next we analyze the perturbation of the $4$-form
potential. Note that
\be C_{0123}=h^{-1}\approx
\frac{r^4}{L^4}\left(1-\frac{2}{5}\frac{b^6}{r^6}+\ldots\right).
\ee
Note that for this class of solutions there is no variation of the 4-form
potential  along the
angular directions. Any dependence on $r$ would violate the Bianchi identity. If fact,
$\calf_5=-r^5\kappa \frac{dh}{dr}
vol(X_5) =L^4\,vol(X_5)$.

We can view the problem in
a slightly more general and useful way. We can perform a reduction
of the ten-dimensional theory on the Sasaki-Einstein space and
identify the masses of the modes that are turned on in our
solution (see appendix \ref{action} for the full details of this
calculation). Schematically, we consider a class of solutions of
the form: \bea
ds_{10}^2&=&L^2[M^2{ds_5}^2+ds_{5'}^2]\\
ds_5^2&=&du^2+e^{2A}dx_{\mu}dx^{\mu}, \qquad
ds_{5'}^2=e^{2B}(d\psi +2\sigma)^2+e^{2C}(e_a^2)\nn \eea
where
$M=-\frac{1}{3}(B+4C)$ and $A, B$ and $C$ are arbitrary functions
of the radial coordinate $u$. The index $a$ runs from 1 to 4 and $e^a$ together form a
four-dimensional {\K}-Einstein metric.  The type IIB solution is
accompanied with a five-form self-dual flux,
\be F_5=\cf +*\cf, \qquad \cf=Q(d\psi +2\s)\w e_1\w e_2\w e_3\w e_4, \ee
where $Q$ is a constant.
The ten-dimensional action we consider is simply \be
S=-\frac{1}{2{\k}^2}\int\left [ d^{10}x \sqrt{g_{10}}R-\frac{1}{2}\cf\w *\cf\right ]. \ee
Introducing the variables $q$ and $f$ defined as $B+4C=\frac{15}{2}q$ and
$B-C=-5f$, after some algebra and expanding the potential for
small values of the modes we obtain
\be S=-\frac{2}{k_5^2}\int d^{5}x \sqrt{g_5}\left[
    \frac{1}{4}\hat{R}-\frac{1}{2}\{15({q'}^2+32q^2)+10({f'}^2+12f^2)\}
    \right]+\ldots
\ee From this effective five-dimensional action we conclude that
the modes corresponding to the class of solutions we consider have
AdS masses equal to $m^2_{f,q}=12, 32$. The modes we found are
well studied in the case of  $T^{1,1}$, they have been presented
in, for example \cite{kt,papa,baryonic}. We have kept basically
the same notation as the original presentation of \cite{kt}
corresponding to $f$ (equation 3.12 in \cite{kt}) and the mass is
determined to be $m_f^2=12$. Using that for scalar excitations
$\Delta=2+\sqrt{4+m^2}$, we conclude that the deformation we are
considering corresponds to a dimension six operator.

The most remarkable result of our calculation is that these modes
are present for any Sasaki-Einstein space. In particular, we see
that $f$ describes squashing of the $U(1)$ fiber with respect to
the K\"ahler-Einstein base. It is worth mentioning that the mode
we are considering is supersymmetric as we have shown explicitly
in appendix \ref{susy}. We, therefore, expect the dual operator to
have protected dimension.

%%%%%%%%%%%%%%%%%%%%%%%%%%%%%%%%%%%%%%%%%%%%%%%%%%%%%%%%%%%%%%%%%%%%%%%%%%%%%%%%%%%%%%%%%%%%%%%%%%%%%%%%%%
\section{A detailed
example: vanishing $\IP^1 \times \IP^1$}\label{kaehler}
%%%%%%%%%%%%%%%%%%%%%%%%%%%%%%%%%%%%%%%%%%%%%%%%%%%%%%%%%%%%%%%%%%%%%%%%%%%%%%%%%%%%%%%%%%%%%%
In this section we consider a metric in which \emph{two} K\"ahler
deformations are present at the same time. Our goal is to describe
the corresponding deformations in the field theory. We also
present an interesting limit yielding the Eguchi-Hanson metric.

Consider the metric found in \cite{pt}

 \bea \label{PZT}
 ds^2&=& \kappa^{-1}(r)dr^2+\frac{1}{9}\kappa(r)r^2(d\psi + \cos\t_1d\phi_1 + \cos\t_2d\phi_2)^2\nonumber \\
&+& \frac{1}{6}r^2 (d\t_1^2 + \sin^2\t_1d\phi_1^2) + \frac{1}{6}
           (r^2 + a^2) (d\t_2^2 + \sin^2\t_2d\phi_2^2),
\eea
where
\be
\kappa(r) = \frac{1 + \frac{9 a^2}{r^2} - \frac{b^6}{r^6}}
{1 + \frac{6  a^2}{r^2}} \,.
\ee

This metric is Calabi-Yau and depends on two real parameters with
dimension of length: $a$ and $b$. It is smooth if the period of
$\psi$ is taken to be $2 \pi$. Algebraically the space can be seen
as the total space of the line bundle ${\cal O}( - K )\rightarrow
\IP^1 \times \IP^1$. The continuous isometries are $SU(2)\times
SU(2) \times U(1)$. The fact that the isometry group has rank $3$
means that the Calabi-Yau in question is toric. In this case
powerful techniques enable us to easily understand algebraically
the geometry. We will use the language of $(p,q)$-webs, developed
in \cite{Aharony:1997bh}. The connection to toric geometry was
derived in \cite{Leung:1997tw}. We view the metric (\ref{PZT}) as
an example of a toric metric on resolved Calabi-Yau cones. In the
general case the isometry is $U(1)^3$, but there are very few
examples where the explicit metric is known.

\begin{figure}[h]
\begin{center}
\epsfig{file=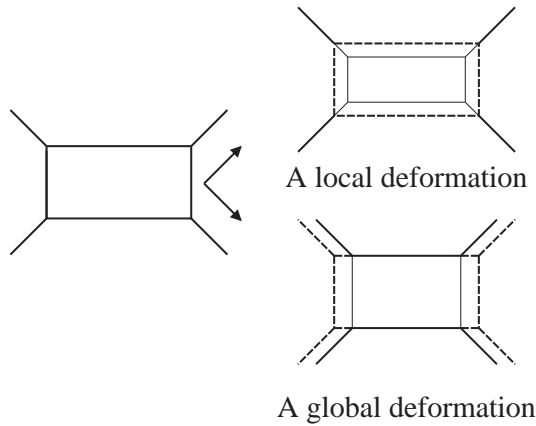,width=7cm} \caption{$(p,q)$ web description of
the vanishing $\IP^1 \times \IP^1$ geometry as in
\cite{Aharony:1997bh}. The two possible deformations are shown as
``global'' and ``local''. \label{pqweb}}
\end{center}
\end{figure}

Toric Calabi-Yau's can be seen as $T^3$ fibrations on a $3$-real
dimensional base. The $(p,q)$ diagram describes the degeneration
locus of this fibration: along the lines a $T^2$ is shrinking to
zero size, and at the cubic vertices the full $T^3$ is vanishing.

The two parameters $a$ and $b$ in (\ref{PZT}) correspond to the
volumes of the two $\IP^1$ in the compact $4$-cycle. More
precisely, $b$ is related to the product of the volumes of the two
$\IP^1$'s, while $a$ is related to the difference between their
areas. In figure \ref{pqweb} the finite segment are $\IP^1$'s, and
the rectangle describes the $4$-cycle $\IP^1 \times \IP^1$.

From the metric (\ref{PZT}) it is easy to see that, asymptotically
at large radius, the $a$ parameter deforms the metric at order
$1/r^2$, while the $b$ deformation only at order $1/r^6$. We can
qualitatively interpret this fact simply looking at the
$(p,q)$-web diagram: the $a$ deformation is changing the position
of the branes at infinity (so it is a \emph{global} deformation),
while the the $b$ deformation does not, so it is a \emph{local}
deformation. In general, global deformations correspond to blowing
up a $2$-cycle, and local deformations correspond to blowing up
$4$-cycles.

In the case of vanishing $b$-deformation, the metric is a $\IZ_2$
orbifold of the resolved conifold, given by (\ref{PZT}) with $b=0$
and the period of $\psi$ taken to be $4\pi$. Notice also that
(\ref{PZT}) at $a=b=0$ is a special example of the
$\mathcal{Y}^{p,q}$ metrics, specifically it is
$\mathcal{Y}^{2,0}$. Equation (\ref{PZT}) gives the most general
K\"ahler Ricci-flat deformation of the conical metric
$\mathcal{Y}^{2,0}$. Notice that the supergravity solution
associated to $\mathcal{Y}^{2,0}$ has also additional moduli. For
instance  there is a moduli space of dual superconformal field
theories (corresponding to $a=b=0$) of complex dimension $5$
\cite{Benvenuti:2005wi}, associated to changing the couplings of
the theory. It is clear that these deformations are present also
for $a$ and $b$ nonzero. It would be nice to investigate these
more general theories.

%%%%%%%%%%%%%%%%%%%%%%%%%%%%%%%%%%%%%%%
\subsection{Limit to Eguchi-Hanson} We now want to show that
a double-scaling limit of (\ref{PZT}) leads to the well-known
Eguchi-Hanson metric. This is expected by looking at the toric
description: sending the area of the rectangle in figure
\ref{pqweb} to infinity while keeping fixed the size of one of the edges, one
gets a $(p,q)$-web with two parallel semi-infinite legs, which is
the space $\IC \times \IC^2/\IZ_2$.

More precisely, sending the parameters $a$ and $b$ to infinity, we
will recover the space $\IC \times ALE$, where $ALE$ in this case
is the total space of the line bundle ${\cal O}(-2) \rightarrow
\IP^1$ (it can also be seen as the K\"ahler Ricci-flat resolution of
$\IC^2/\IZ_2$).  We take the limit $(a, b) \rightarrow \infty$
with
\be
 c^4 = \frac{4 b^6}{81 a^2}
\ee fixed. In other words we focus on region of the geometry where
$r\ll a,b$ but $r$ is of the same order as $c$.
%(taking $r<<c$ as well one would get $\IC^3$)
 In this limit $\kappa(r)$ becomes
 \be
\kappa(r) = \frac{\frac{r^2}{a^2} + 9 - \frac{81 c^4}{4 r^4}}{\frac{r^2}{a^2} + 6}
 \rightarrow \frac{3}{2} (1 - \frac{9 c^4}{4 r^4}).
 \ee
Let us now consider the term in \ref{PZT} proportional to
$(r^2+a^2)$. In order to find a finite limit we rescale the
coordinates $\t_2$ and $\phi_2$ by a factor of $a$:
$\tilde\t=\frac{\t_2}{a}$, $\tilde{\phi}=\frac{\phi_2}{a}$,
 \be \label{flatC}
  (r^2+a^2)(d\t_2^2 + \sin^2\t_2d\phi_2^2) \rightarrow d\tilde{\t}^2 + \tilde{\t}^2
  d\tilde{\phi}^2= d\textrm{x}^2_2.
 \ee
We thus get the flat metric on $\IC$, as expected from the fact
that we are simply zooming in on a point of a smooth $2$-sphere.
The piece in (\ref{PZT}) proportional to $\kappa(r)$, keeping
track of the rescaling of $\phi_2$ becomes
 \be
  d\psi + \cos\t_1d\phi_1 + \cos\t_2d\phi_2 \rightarrow d\psi +
  \cos\t_1d\phi_1.
 \ee
At the end we see that the metric decomposes in one on $\IC$
(\ref{flatC}) and a remaining non-compact $4$-manifold: rescaling
$R=r \sqrt{2/3}$
 \be\label{EH}
  ds^2 =  \frac{dR^2}{1 - \frac{c^4}{R^4}} +
      \frac{1}{4} (1 - \frac{c^4}{R^4}) R^2 (d\psi + \cos\t_1d\phi_1)^2
   +  \frac{1}{4} R^2 (d\t_1^2 + \sin^2\t_1d\phi_1^2) + d\textrm{x}^2_2.
 \ee
This is precisely the well-known Eguchi-Hanson metric. From a
three dimensional point of view the limit space has isometries
$SU(2) \times U(1) \times (\IR^2 \times SO(2))$.

\subsection{The field theory dual} In this section we discuss the
gauge theory living on $N$ D3 branes probing the local $\IP^1
\times \IP^1$. The quiver diagram is well known: it has $4$ nodes
and is reproduced in figure \ref{pqweb}. The superpotential is
readily obtained by imposing $SU(2) \times SU(2)$ invariance or by
orbifolding the conifold field theory:
 \be
 \label{sp}
 \mathcal{W} = \l tr \left(A_{\a} B_{\ad} C_{\b} D_{\bd} \right) \varepsilon^{\a \, \b} \varepsilon^{\ad \, \bd}.
 \ee

\begin{figure}[h]
\begin{center}
\epsfig{file=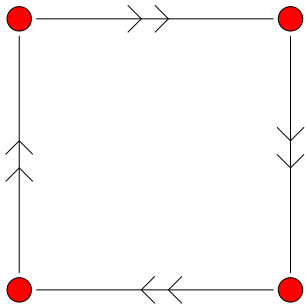,width=4cm} \caption{Quiver diagram for D3 branes
probing the local $\IP^1 \times \IP^1$ singularity. \label{quiver}}
\end{center}
\end{figure}

We are interested in the moduli space of vacua of this field
theory, which we obtain by analyzing $\mathcal{F}$ and
$\mathcal{D}$ term relations on the fundamental fields of the
theory. The full moduli space of the theory has complex dimension
$3N$. $3 ( N - 1 )$ flat directions correspond to motion of the
$N$ D3 branes in the Calabi-Yau. These are what we call ``mesonic
flat directions.'' The remaining $3$ flat directions will be seen
to be associated to the deformations of the geometry or to turning
on $B$-fields. These can be called ``non-mesonic'' vevs. We will
make use of the terminology of Fayet-Ilioupoulos parameters in
order to describe the non-mesonic directions, even if this is not
completely correct.

For our purposes it is enough to consider the case that the $N
\times N$ matrices $A_{\a}, B_{\ad}, C_{\b}, D_{\bd}$ commute. For
the moment we also assume that all these $8$ complex matrices are
proportional to the identity, corresponding to keeping the $N$ D3
branes coincident.\footnote{This is equivalent to considering the
Calabi-Yau only one time instead of the $N$-fold symmetrized
product of it. It will be simple to relax this assumption in the
following,  when we will consider distributions of D3 branes
smeared on the blown up cycles.} It is easy to see that the
$\mathcal{F}$ terms are solved if and only if
 \bea
 A_{\a} & \propto C_{\a},\\
 B_{\ad} & \propto D_{\ad}\,,
 \eea
which is equivalent to
 \bea
 A_{\a}  = A \, e^{i\t_A} \, n_{\a} &  C_{\a}  = C \, e^{i\t_C} \, n_{\a}\\
 B_{\ad}  = B \, e^{i\t_B} \, m_{\ad} & D_{\ad}  = D \, e^{i\t_D} \, m_{\ad}\,,
 \eea
where $A, B, C, D$ are $4$ \emph{positive} real numbers, the $\t$'s
are phases, $n_{\a}$ is parameterizing a $\IP^1$ of unit radius
and $m_{\ad}$ is parameterizing another $\IP^1$ of unit radius. We
already see the emergence of the geometry ($\IP^1 \times \IP^1$).
 We now quotient by the gauge groups. These act by shifting the phases $\t$. Using this
 freedom all the phases can be taken to be equal to $\psi$. $\mathcal{D}$
 term relations read
  \bea \label{FI}
 \nonumber A^2 - B^2 = t_1,\\
  B^2 - C^2 = t_2,\\
 \nonumber C^2 - D^2 = t_3,\\
 \nonumber D^2 - A^2 = t_4,
  \eea
where $t_i$ are the $4$ Fayet-Iliopoulos terms,\footnote{Strictly
speaking, after taking the near horizon limit, the gauge group is
$SU(N)^4$, not $U(N)^4$, so no FI parameters can be turned on. It
is possible to repeat our discussion just in terms of the vev of
the fields. There are $3(N-1)$ complex directions corresponding to
giving vev to mesonic operators. There are moreover $3$ flat
directions corresponding, formally, to the $3$ FI terms we are
discussing. The $3$ relative phases $\t$s combine with these to
give complex flat directions.} that satisfy $t_1+t_2+t_3+t_4=0$,
since in the sum of the $4$ previous relations the l.h.s. vanishes
identically. This can also be seen by noticing that the diagonal
$U(1)$ inside $U(N)^4$ is decoupled, since all fields are
uncharged under it.

Notice that the solutions of eqs (\ref{FI}) are parameterized by a
positive real line, and that the ``smallest'' possible vev is
obtained when at least one of $A, B, C, D$ vanishes. Giving vev
where one of these vanishes correspond to keeping the D3 branes at
the smallest possible radius ($r=b$), and mesonic chiral operators
are not taking vevs. If all the $A, B, C, D$ are non-vanishing,
then also mesonic operators are taking vevs; this corresponds to
moving the branes outside the resolved cycle.

We thus see that the mesonic moduli space of vacua is
parameterized by $6$ real coordinates: $n_{\a}$, $m_{\ad}$, $\psi$
and one ``radial'' direction coming from $A, B, C, D$. It is
simple to match these coordinates with the coordinates $(\t_1,
\p_1)$, $(\t_2, \p_2)$, $\psi$, $r$.

We are now interested in an explicit map between the FI parameters
$t_i$ in the gauge theory and the deformation parameters $a$ and
$b$ in the geometry. Clearly the ``conical'' geometry $a=b=0$
corresponds to vanishing FI parameters, i.e. to conformal
invariance.
%\begin{figure}\includegraphics[height=4.5cm]{F0.eps}
% \caption{Quiver theory for D3-branes at a point in the local$\IP^1 \times \IP^1$ Calabi-Yau. \label{figure}}
%\end{figure}

%%%%%%%%%%%%%%%%%%%%%%%%%%%%%%%%%%%%%%%%%%%%%%%%%%%%%%%%%%%%%%%%%%%%
\subsection{The global deformation} Turning on only the global
``$a$ deformation,'' the space is the $\IZ_2$ orbifold of the
resolved conifold, so also in the field theory, we just have to
``orbifold'' the analysis of \cite{kw1}. The final result is
 \be\label{FIa}
 t_1 = - t_2 = t_3 = -t_4  \propto a^2 > 0
 \ee
Let us check this claim directly. Eqs. (\ref{FI}) in this case
admit a ``minimal'' solution with
 \be
 A = C = a_f \,, \;\;\;\;\;\; B = D =0
 \ee
We thus see that the FI (\ref{FIa}) are giving finite volume to
the $\IP^1$ parameterized by the $n_{\a}$, in accordance with the
metric (\ref{PZT}). At energies below the scale $t$, some fields
are eaten by the Higgs mechanism and the field theory reduces in
the infrared to a different field theory. The $U(N)$ gauge
symmetries associated to node $1$ and $4$ are broken into a
diagonal $U(N)$, because of the vev taken by $A$. Similar is the
case for nodes $2$ and $3$. The infrared gauge symmetry is thus
$U(N) \times U(N)$.

Let us assume that $A_1= C_1$ are the fields taking the vev $a_f$,
that combine with the broken $U(N) \times U(N)$ gauge fields into
massive gauge bosons. This assumption corresponds to putting the
D3 branes at the ``north-pole'' of the blown-up $\IP^1$ (of course
any other point would lead to the same discussion, by $SU(2)$
symmetry). From the geometry, it is clear that the $a \rightarrow
\infty$ limit, focusing on this point, is given by $\IC \times
\IC^2/\IZ_2$.

The IR quiver has only $6$ chiral fields. $A_2$ and $C_2$
transform as adjoint fields and the other $4$ fields as
bifundamentals. Because of the $\varepsilon^{\a \b}$ in the
superpotential (\ref{sp}), we see that no massive terms in the
superpotential are generated. All the superpotential terms are
cubic and proportional to $\l \, \a_f$:
 \be
 \mathcal{W}_{IR} =  a_f\, \l \; tr \left( A_2 B_{\ad} D_{\bd} - B_{\ad} C_2 D_{\bd}
 \right) \varepsilon^{\ad \bd}
 \ee
This is precisely the matter content and superpotential
corresponding the $\mathcal{N}=2$ field theory living at $\IC
\times \IC^2/\IZ_2$.

Let us end this paragraph studying the baryonic current as for the
almost identical case of the conifold \cite{kw1}:
 \be
 \mathcal{K}_B = A_{\a} \bar{A}^{\a} - B_{\ad} \bar{B}^{\ad} + C_{\a} \bar{C}^{\a} - D_{\ad} \bar{D}^{\ad}.
 \ee
We see that
 \be
 <\mathcal{K}_B> \propto a^2.
 \ee
This is precisely the dimension $2$ operator corresponding to the
harmonic turned on at first order in the geometry, which at large
$r$ undergoes corrections of order $1/r^2$. This order parameter
has protected dimension since it is a conserved multiplet. The
harmonic has its origin in the so called Betti multiplets, and was
identified (for the very similar case of the conifold) in
\cite{t11spec}.

%%%%%%%%%%%%%%%%%%%%%%%%%%%%%%%%%%%%%%%%%%%%%%%%%%%%%%%%%%%%%%%%%%%%%%
\subsubsection{Global deformations from the metric: resolved conifold}
%%%%%%%%%%%%%%%%%%%%%%%%%%%%%%%%%%%%%%%%%%%%%%%%%%%%%%%%%%%%%%%%%%%%%%
The analysis presented in the previous subsections is nothing but an
extension of the description of the small resolution of the conifold. Let us briefly recall what the
precise picture is as presented by Klebanov and Witten in \cite{kw1}. They
argued that the natural gauge theory order parameter would be the
lowest component of the baryonic supercurrent:
 \be
  \calu=\Tr (A_i \bar{A}_i-\bar{B}_j B_j).
 \ee
$\calu$ has dimension two.  A detailed analysis of this operator
including its origin in the Betti multiplet is given in
\cite{t11spec}. Given the explicit solution of D3 branes on the
resolved conifold, we can quantitatively verify this claim.
Namely, the supergravity solution in question is of the form
(\ref{d3}) with \cite{resolved}: \bea ds_6^2&=&\kappa(r)^{-1}dr^2
+ \kappa \, r^2 e_{\psi}^2 + r^2 (e_{\theta_1}^2+e_{\phi_1}^2)
+ (r^2+6a^2) (e_{\theta_1}^2+e_{\phi_1}^2), \nonumber \\
h&=&\frac{2L^4}{9a^2 r^2}-\frac{2L^4}{81
a^4}\ln\left(1+\frac{9a^2}{r^2}\right), \qquad
\kappa=\frac{r^2+9a^2}{r^2+6a^2}. \eea The linearization of the
solution can be written as: \bea
ds^2&=&\frac{r^2}{L^2}dx_\mu dx^\mu + \frac{L^2}{r^2} dr^2 + L^2 ds^2(T^{1,1}) \nonumber \\
&+& \frac{3a^2}{L^2} dx_\mu dx_\mu - \frac{6L^2a^2}{r^4}dr^2 - \frac{3L^2a^2}{r^2} ds^2(X_4) \nonumber \\
&+&\frac{36 L^2a^4}{r^4}(d\psi +\sigma)^2+ \ldots \eea In the
first line, we have simply $AdS_5\times T^{1,1}$. From the second
line we read that the solution corresponds to give a vacuum
expectation value to a dimension two operator. One naturally has
\be <\calu>\approx a^2. \ee

The description above can be understood from the quiver gauge
theory point of view. A convenient way to describe the small
resolution of the conifold is in terms of four complex numbers
satisfying the real constraint \cite{kw1} \be
|a_1|^2+|a_2|^2-|b_1|^2-|b_2|^2=t, \ee where t is the area of the
$P^1$, and then one takes the quotient by a $U(1)$ action. This is
usually said as a gauged linear sigma model with $4$ fields, a
$U(1)$ gauge group, where the charges of the fields are
$(1,1,-1,-1)$. The above relation makes it clear that from the
four dimensional quiver gauge theory point of view one is
introducing a Fayet-Iliopoulos parameter.

%%%%%%%%%%%%%%%%%%%%%%%%%%%%%%%%%%%%%%%%%%%%%%%
\subsection{General geometric deformation}
We are left with other two FI parameters, and one other geometric
deformation, that we called $b$. In the field theory there are two
parameters, instead of one, because the field theory sees the full
stringy moduli space, and the latter also includes a $B$-field
that can be turned on.

Instead of talking in terms of FI parameters,\footnote{The global
deformations are in correspondence with baryonic symmetries in the
gauge theories. Gauging these symmetries one can talk about FI
parameters. The other FI parameters instead correspond to
anomalous $U(1)$ transformations in the gauge theory, so they
cannot be gauged.} we exhibit a possible solution on the moduli
space of vacua of the gauge theory which corresponds to turning on
both the $a$ and $b$ deformation. Consider giving vevs of the form
 \bea\label{vevb}
 <A_{\a}> & = &\d_{1 \a} \, ( a_f + b_f )\\
 <B_{\ad}> & = &\d_{1 \ad} \, \sqrt{2} b_f  \\
 <C_{\a}> & = &\d_{1 \a} \, ( a_f - b_f )\\
 <D_{\ad}> & = &0
 \eea
Notice that the vev of the baryonic current is proportional to
$a_f^2$ and does not depend on $b_f$. At energies greater than
both $a_f$ and $b_f$ we have the UV conformal field theory with
gauge symmetry $U(N)^4$. At energies $E$ such that $b_f < E <
a_f$, we have the $\mathcal{N}=2$ field theory discussed above.
Now, however, one bifundamental field, $B_{\dot{1}}$ is taking a
vev $\sqrt{2} b_f$.

At energies $E < b_f$ the field theory is Higgsed to a $U(N)$
field theory with $5$ adjoints: $A_2, C_2, B_{\dot{2}},
D_{\dot{1}}, D_{\dot{2}}$. The superpotential
 \be
 W =  \l \; tr \left( (a_f - b_f) b_f A_2 D_{2} - (a_f - b_f) A_2 B_{2} D_{1} + (a_f + b_f) B_{2} C_2 D_{1} - (a_f + b_f) b_f C_2 D_{2}
 \right)
 \ee
contains mass terms for the fields $D_2$ and $(a_f - b_f)A_2-(a_f
+ b_f)C_2$; integrating these out, one finds the $\mathcal{N}=4$
SYM.

The conclusion is that the vevs (\ref{vevb}) are in correspondence
with the two geometric deformations discussed in the previous
paragraph.

\subsubsection*{Comments on smearing of the D3 branes}
For the gauge theory discussed above (all the D3 branes coincident
and localized), the SUGRA metric including the gravitational
backreaction of the D3 branes would break the $SU(2) \times SU(2)$
symmetry of rotations of the $n_{\a}$ $\IP^1$, and for instance
the warp factor would not be of the form (\ref{warp}). Putting
instead a uniform distribution of D3 branes on the blown up $\IP^1
\times \IP^1$, one can find a nice explicit supergravity solution,
which is expected to be dual to the non-conformal gauge theory
living on this distribution of D3 branes. In this case the
solution is written in term of the Calabi-Yau metric as in
(\ref{d3}) and the warp factor is like (\ref{warp}), depending
only on $r$.

One general lesson from this discussion is that at order $1/r^2$
the corrections to the conformal solutions correspond to blown up
$2$-cycles. Keeping these ``global'' deformations fixed, the order
$1/r^4$ deformation only sees the total number of D3 branes. We
can instead expect that corrections coming from how we distribute
the branes on the blown up $4$-cycle play an important role at
higher orders, starting at $1/r^6$. This should be analogous to
the Coulomb branch solutions for $\mathcal{N}=4$ SYM \cite{kw1}.

%%%%%%%%%%%%%%%%%%%%%%%%%%%%%%%%%%%%%%%%%%%%%%%%%%%%%%%%%%
\subsection{Toric description of the blow up}\label{blowupexample}
Let us describe the process of the blowing-up from the
point of view of toric geometry.
\begin{figure}\begin{center}
\epsfig{file=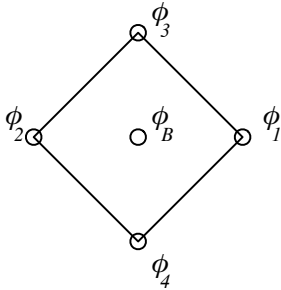,width=4cm}
\end{center}
\caption{Toric description of the four-cycle in the orbifolded conifold.
\label{toric-p1p1}}
\end{figure}
The toric data before the blow up consist of
four points on the outer square in figure \ref{toric-p1p1}.
In the language of two-dimensional gauged linear sigma model,
we introduce chiral superfields $\phi_{1,2,3,4}$
for each of the toric datum,
and we introduce a vector multiplet $V_1$ to kill
extra degrees of  freedom.
The toric data and the charge assignment is
tabulated in the left hand side of table \ref{chargetable}.
\begin{table}\[
\begin{array}{c|cccc}
&\phi_1&\phi_2&\phi_3&\phi_4\\
\hline
&1&1&1&1 \\
&1&-1&0&0\\
&0&0&1&-1\\
V_1& 1&1&-1&-1
\end{array}
\Longrightarrow
\begin{array}{c|ccccc}
&\phi_1&\phi_2&\phi_3&\phi_4&\phi_B\\
\hline
&1&1&1&1 &1\\
&1&-1&0&0&0\\
&0&0&1&-1&0\\
V_1& 1&1&-1&-1&0\\
V_2&1&1&1&1&-4
\end{array}
\]
\caption{The charge assignments\label{chargetable}}
\end{table}

Thus the toric manifold is described by
the equations \begin{equation}
| \phi_1|^2+| \phi_2|^2-| \phi_3|^2-| \phi_4|^2 = 0
\end{equation} divided by the identification
\begin{equation}
(\phi_1,\phi_2,\phi_3,\phi_4)\sim
(e^{i\theta}\phi_1,e^{i\theta}\phi_2,e^{-i\theta}\phi_3,e^{-i\theta}\phi_4).
\end{equation}
This is precisely the cone over orbifolded $T^{1,1}$.

In order to blow up the four-cycle corresponding to the internal
point in figure \ref{toric-p1p1}, we add a corresponding chiral superfield
$\phi_B$ and another vector superfield as in the right hand side of the
table \ref{chargetable}. Now  the equations describing the
toric manifold are
\begin{eqnarray}
| \phi_1|^2+| \phi_2|^2-| \phi_3|^2-| \phi_4|^2 &=& 0\label{firsteq}\\
| \phi_1|^2+| \phi_2|^2+| \phi_3|^2+| \phi_4|^2-4|\phi_B|^2 &=& t
\end{eqnarray}
under the identification
\begin{eqnarray}
(\phi_1,\phi_2,\phi_3,\phi_4,\phi_B)&\sim&
(e^{i\theta}\phi_1,e^{i\theta}\phi_2,e^{-i\theta}\phi_3,e^{-i\theta}\phi_4
,\phi_B),\label{ident1}\\
(\phi_1,\phi_2,\phi_3,\phi_4,\phi_B)&\sim&
(e^{i\theta}\phi_1,e^{i\theta}\phi_2,e^{i\theta}\phi_3,e^{i\theta}\phi_4
,e^{-4i\theta}\phi_B).\label{ident2}
\end{eqnarray}where $t\ge 0$ is the Fayet-Iliopoulos term
for the vector multiplet $V_2$, which controls
the size of the blown-up four-cycle.

Let us explicitly see that we have $\mathbb{P}^1\times \mathbb{P}^1$ of non-zero size.
Let us slice the manifold at constant $|\phi_B|^2=c$.
If $c>0$, we can fix the second identification (\ref{ident2})
by setting $\phi_B$ at a positive real number.
The rest of the equation is
\begin{equation}
|\phi_1|^2+|\phi_2|^2=|\phi_3|^2+|\phi_4|^2=t/2+2c,\label{S3xS3}
\end{equation}
which is a product of two $S^3$ of the same size.
We get $T^{1,1}/Z_2$ after dividing by the identification (\ref{ident1}).
When $c=0$,  we can no longer fix (\ref{ident2}) using $\phi_B$.
Thus, we need to divide (\ref{S3xS3}) by (\ref{ident1}) and (\ref{ident2}),
which yields $\mathbb{P}^1\times \mathbb{P}^1$ of radius $\sqrt{t/2}$.
This is precisely the local deformation discussed in the previous
subsections.

We can also introduce the Fayet-Iliopoulos term to (\ref{firsteq}).
One can easily see that it changes the relative size of two $P_1$.
It is the global deformation in our parlance.

%%%%%%%%%%%%%%%%%%%%%%%%%%%%%%%%%%%%%%%%%%%%%%
\subsection{Comments on the 4-cycle deformation of the Klebanov-Strassler solution}

In this section we have seen that a $\IZ_2$ orbifold of the
conifold admits $3$ deformations associated to giving vev to
fundamental fields (these are $3$ non mesonic direction in the
field theory moduli space). One deformation corresponds to a blown
up $2$ cycle, one to a blown up $4$ cycle, one to a $B$ field. In
the case of the conifold only one deformation is present, the so
called resolved conifold, corresponding to a blown up $2$ cycle,
which can be called the ``baryonic flat direction.''

We know that the resolution of the conifold plays an important
role also after the addition of fractional branes. The moduli
space of vacua (for certain values of the number of fractional
branes) contains a baryonic branch, which comes from the ``baryonic
flat direction'' discussed above. For a thorough analysis see
\cite{Dymarsky:2005xt}. The Klebanov-Strassler \cite{ks} solution
corresponds to a field theory sitting on this baryonic branch at
the special point $a=0$ (in our notation).

Very interestingly, there is indeed a generalization of the
Klebanov-Strassler background corresponding to this `` K\"ahler''
modulus, this is the so called baryonic branch solution, found for
small deformation parameters in \cite{Gubser:2004qj} and at all
orders, using $SU(3)$ structures, in \cite{baryonic}.

From our analysis it is natural to expect that also for the local
$\IP^1 \times \IP^1$ geometry, after adding fractional branes,
there is a non-mesonic branch. It is very simple to obtain the
baryonic branch solution with fractional branes for our case: we
just have to take the solution of \cite{baryonic} but declare the
periodicity of the angle $\psi$ to be $2\pi$ instead of $4\pi$.

The point is that this $\IZ_2$ orbifold of the baryonic branch
also admits a deformation corresponding to the $b$ parameter. We
think it would be very interesting to study this problem. Notice
that the full solution should be captured by functions of only one
variables (as in \cite{baryonic}), since the $b$ deformation is
not breaking any further continuous symmetry. As a first step it
would be interesting to identify the massless mode that should
appear in the twisted sector of the orbifold. It would also be
important to improve the field theory analysis along the lines of
\cite{Dymarsky:2005xt}.
%%%%%%%%%%%%%%%%%%%%%%%%%%%%%%%%%%%%%%%%%%%%%%%%%%%%

%%%%%%%%%%%%%%%%%%%%%%%%%%%%%%%%%%%%%%%%%%%%%%%
\section{Blowing up $4$-cycles: the general case}\label{general}
%%%%%%%%%%%%%%%%%%%%%%%%%%%%%%%%%%%%%%%%%%%%%%

%%%%%%%%%%%%%%%%%%%%%%%%%%%%%%%%%%%%%%
\subsection{Deformations and Operators}
%The $(p,q)$-web approach
%%%%%%%%%%%%%%%%%%%%%%%%%%%%%%%%%%%%%%%%%%
Let us consider various aspects of the general case. Given a
resolved local Calabi-Yau, there a two types of K\"ahler
deformations. One corresponds to changing the size of a $\mathbb{P}^1$ as
described by the parameter $a$, the other to changing the size of
a four cycle described by the $b$-parameter.
For toric CY there is
a dual description in terms of a $(p,q)$-web. The $a$-deformation
changes the position of the branes at infinity, while the
$b$-deformation is a ``local'' deformation, it only change the sizes
of the internal faces, without moving the external branes. In this
sense the $a$-deformation is heavy and requires a lot of energy,
while the local $b$-deformation
 does not require a lot of energy.
For a complicated CY there are many independent $a$- and $b$-deformations.
The number of $a$-deformations is the
number of external legs (denoted by  $d$ ) minus 3.
The number of $b$-deformations is the number of internal points in
the toric diagram (denoted by $I$ ), which will be discussed
further in section \ref{globalissues}.
The universal $b$-deformation we described in section \ref{d3branes}
is a particular example of this class.
%the one that gives the same volume to all the compact 4-cycles.

From the examples considered in this paper it is natural to
conjecture that the change of the metric at large radius is always
of the type $1/r^2$ for $a$-deformations, while it is $1/r^6$ for
$b$-deformations.

The $a$-deformation in the field theory has  as order parameter
the corresponding (scalar superpartner of the) baryonic current,
which has always dimension $2$, matching with the $1/r^2$
correction to the metric. The number of global deformation is
always given by the number of external legs in the $(p,q)$ diagram
minus $3$.

For the $b$-deformation we expect to find as many dimension six
operators in the field theory as four cycles in the geometry which
coincide with the number of internal points in the toric diagram.
The total number of gauge groups is always given by twice the area
of the toric diagram. The area, by Pick's theorem, is expressible
in terms of the number of external points $d$ and the number of
internal points $I$. Concluding, the number of gauge groups is
 \be
 \#\textrm{gauge-groups}= 1 + (d-3) + 2 I.
 \ee

Now, the number of formal FI parameters that can be turned on is
the number of gauge groups minus $1$
(recall no field is charged under the diagonal $U(1)$).
$d-3$ such FI parameters are associated to the $d-3$
baryonic symmetries. Among the other $2 I$ FI parameters,
$I$ are associated to the $I$ $4$-cycles that can be blown up.
The remaining $I$ FI parameters have thus to be associated to
$B$-fields that can be turned on in the supergravity solution.

We have thus matched the number of the three possible deformations
(global, local and $B$-fields) with the formal FI parameters in the
field theory.

This matching strongly suggests that the operators of dimension
six turned on by the local deformation are associated with the
gauge groups in the quiver. We thus propose that these operators
are roughly of the form:

\be
 \mathcal{O}_i = \sum_{g} c_{i, g} \mathcal{W}_g\bar{\mathcal{W}_g}
\ee
where the sum is over the gauge groups in the quiver.

Notice that similar operators are present in the supergravity
background of $T^{1,1}$ \cite{gubser, t11spec}, and are also
turned on in the baryonic branch of the Klebanov-Strassler solution
\cite{baryonic}.

\subsection{More on $a$-deformation}
\begin{figure}
\centerline{\epsfig{file=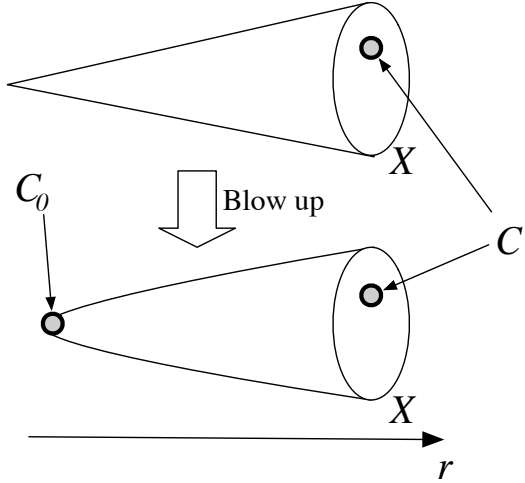,width=7cm}}
\caption{$a$-type deformation \label{a-deform}}
\end{figure}

Indeed, we can check that the dimension of
operators corresponding to the $a$-deformations
are two, and
there are $d-3$ of them for the compactification
on the toric Sasaki-Einstein $X$. %, where $d$ is the number of
%external points of the toric diagram for $X$.
The argument is a generalization of one given in
\cite{kw1}.
The original metric on the cone is \begin{equation}
ds^2=dr^2+r^2 ds_X^2.
\end{equation}
Let us blow up some two-cycles and four-cycles
at the tip of the cone (see figure \ref{a-deform}),
and let us denote the K\"ahler form before and after the blowup
by $J$ and $J+\delta J$, respectively.
Note that $J$ behaves as
\begin{equation}
J\sim r^2 \label{Jbehavior}
\end{equation}
when $r\to \infty$.

Suppose that there is a supersymmetric two-cycle $C_0$
at the tip which comes from a two-cycle $C$ in the Sasaki-Einstein $X$,
and let us blow-up so that $C_0$ has area $t$.
Then,
\begin{equation}
t=\int_{C_0} \sqrt{g} = \int_{C_0} J+\delta J,
\end{equation}
because the supersymmetric cycle should be
calibrated with respect to the K\"ahler form. Now, since $C_0$ and $C$
are in the same homology class, we have
\begin{equation}
t=\int_C J+\delta J.
\end{equation}
Since we assumed that $C$ is a nontrivial homology class in $H_2(X)$,
we can take $C$ at arbitrarily large $r$.
It means that $\delta J$ has the behavior $\sim r^0$
under $r\to \infty$. Compared to (\ref{Jbehavior}),
it is a $1/r^2$ type deformation, which corresponds to a
dimension two operator in the dual CFT.
Thus we see that there is an $a$-type deformation
for each homology class in $H_2(X)$.
From Poincar\'e duality in $X$, $H_2(X)$ has the same
dimension $d-3$
as $H_3(X)$, which corresponds to the baryonic symmetries.

From the viewpoint of the SCFT,
the superconformal algebra fixes the dimension of
the scalar operator in the superconformal multiplet
of a conserved flavor current so that it has dimension two.
For the SCFT dual to the type IIB theory on $AdS_5 \times X$,
there is $d-1$ flavor symmetry.
$d-3$ of them are the so-called baryonic symmetries,
and the modes we found for each of the homology class
in $H_2(X)$ are the dual manifestation in the gravity side.
It would be interesting to find the remaining two modes.
This description allows us to canonically identify corresponding operator as the lowest component of the
supercurrent that generates the baryonic symmetry.

\subsection{Global issues}\label{globalissues}
Next, let us discuss the issues concerning the
global topology of the $b$-deformation.
Recall that in section \ref{d3branes} we obtained that near the
point $r\approx b$, the metric takes the form:
\be \label{4cycle}
ds^2\approx
du^2+9u^2 (d\psi+\sigma)^2 + b^2 ds_4^2.
\ee
In order to avoid
conical singularities, we require the periodicity of $3\psi$ to be
$2\pi$. In the case of the conifold, what we call $3\psi$ here,
has period $4\pi$. In order to avoid conical singularities we
must, therefore, introduce a $\mathbb{Z}_2$ orbifold\footnote{The
need to perform this quotient was mentioned in \cite{pt}, here we
see that it is a particular case of a more general situation.}
ending up with $T^{1,1}/\mathbb{Z}_2$. Similarly,  for $S^5$
written as a $U(1)$ bundle over $\mathbb{C}\mathbb{P}^2$, we have
to impose a $\mathbb{Z}_3$ orbifold in order to avoid conical
singularities and we are left with $S^5/\mathbb{Z}_3$. For the
general $Y^{p,q}$ and $L^{p,q,r}$, in the quasi-regular case a specific
quotient is required.

\begin{figure}[h]
\begin{center}
\epsfig{file=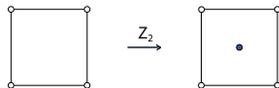,width=4cm} \caption{A four cycle in the
orbifolded conifold. \label{conz2}}
\end{center}
\end{figure}

There is a beautiful picture of the above situation that arises
from the toric diagram of the original Calabi-Yau singularity. As
can be seen from the metric in (\ref{4cycle}) at the origin of
space $r=b$, there is a 4-cycle with finite volume proportional to
$b$. We can try to view this 4-cycle more algebraically. A
particularly useful tool is the toric diagram. Let us consider the
example of the conifold. Recall that 4-cycles that can be blown up
appear in the toric diagram as internal points. Therefore, we
expect to relate the quotient to the generation of internal points
in the toric diagram. Let us take the example of the conifold. The
toric diagram of the conifold is simply a square with no internal
points. After the quotient by $\mathbb{Z}_2$, we have precisely an
internal point  which we show in blue in figure \ref{conz2}.
We think of the orbifold in the sense of \be \mathbb{Z}^2
\subseteq L \subseteq \left(\frac12\mathbb{Z}\right)^2, \ee where
\be L:=\{(n_1,n_2):\,\, \sum n_i \in \mathbb{Z}\}. \ee Clearly the
point $(\frac12,\frac12)$ belongs to the lattice $L$  and
corresponds to the point in the middle in figure \ref{conz2}.
A concrete process of blowing up a four-cycle
corresponding to an internal point
is already presented in section \ref{blowupexample}.

\begin{figure}[!h]
\begin{center}
\epsfig{file=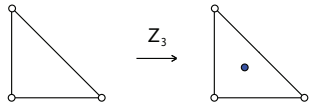,width=4cm} \caption{A four cycle in the
orbifolded $\mathbb{C}^3$. \label{c3z3}}
\end{center}
\end{figure}

For $\mathbb{C}^3$, after a $\mathbb{Z}_3$ orbifold, we see that
the toric diagram contains an internal point, also shown in blue
in figure \ref{c3z3}.

The situation for the $\mathbb{Z}_3$ orbifold of $\mathbb{C}^3$ is
conceptually similar although technically slightly more
complicated. The off shot is that the quotient indeed generates an
internal point. Basically, taking the quotient
$\mathbb{C}^3/\mathbb{Z}_3$ adds a point with coordinates
$\frac13(1,1,1)$ in a lattice where the original triangle had
coordinates $(1,0,0), (0,1,0)$ and $(0,0,1)$. This point
corresponds precisely to acting with a cube root of one contained
in $(\mathbb{C}^*)^3$. The internal point algebraically describes
a blow up mode which is geometrically captured in equation
\ref{4cycle}.

%%%%%%%%%%%%%%%%%%%%%%%%%%%%%%%%%%%%%%%%%%%%%%%%%%%%%
%\subsection{Toric description of $b$-deformation}
%%%%%%%%%%%%%%%%%%%%%%%%%%%%%%%%%%%%%%%%%%%%%%%%%
%we give a description of the K\"ahler $b$-deformation in terms
%of toric geometry.
The observation above can be extended to any of the internal
points of the toric diagram.
We follow the notation and language of \cite{fulton}.
Let us choose the basis of the $N$ lattice such that
the toric data  can be given as $k_I=(1,\vec k_I)$.
When the Sasaki-Einstein is regular, then the Reeb vector $R$
is the integral combination of the generator of $T^3$
defining the toric geometry. It means $R$ is on the $N$ lattice.
Furthermore, it is guaranteed to be on the same plane in $N$
as the other toric data. Thus $R=(1,\vec R)$ with integer entries.

Now add $R$ to the toric data $\{k_1,k_2,\ldots\}$
and consider the toric manifold $Z$ defined by these points.
$Z$ is a blowup of the cone over $X$, and since all the toric data
$R$ and $k_I$ lie on a plane, $Z$ is at least topologically Calabi-Yau.
Let us recall that, in the gauged linear sigma model description, we introduce
chiral superfields $x_v$ for each $v$ in the toric data
and do the symplectic reduction.
Thus  for each vector $v$ in the toric data,
we have a divisor $D_v$ defined by $x_v=0$
where the vector $v$ degenerates.
We can now see that $Z$ has precisely the topology
of the $b$-deformation of the cone over $X$.

%%%%%%%%%%%%%%%%%%%%%%%%%%%%%%%%%%%%%%%%%%%%%%%%%%%%%%%%%%%%%%
\subsection{Topological restriction of the $b$-deformation}
%%%%%%%%%%%%%%%%%%%%%%%%%%%%%%%%%%%%%%%%%%%%%%%%%%%
Let us elaborate the discussion on the need for the orbifolding.
First let us see, in order for the $b$-deformation to make sense,
the Sasaki-Einstein manifold needs to be regular or quasi-regular.
It can be argued in many ways. One way goes as follows:
near $r\sim b$, the metric looks locally like \eqref{4cycle}.
%
%\begin{equation}
%du^2 + u^2( d\psi + \sigma)^2 + ds_{KE}^2.
%\label{semfld}
%\end{equation}
%Here $\partial_\psi $ is the Reeb vector.
Fix $u=u_0$ small and nonzero. Then the $u=u_0$ surface is
topologically the same as the original Sasaki-Einstein $X$.
Suppose the base K\"ahler-Einstein manifold $Y$ exist globally.
Forgetting the $\psi$ coordinate determines the map $p:X\to Y$.
If two points $a$ and $b$ on $X$ are related by the action of $\partial_\psi$
then $p(a)=p(b)$.   In the regular or quasi-regular case,
a circle on $X$ projects to a point on $Y$.
However in the irregular case, the action of the Reeb vector $\partial_\psi$
does not close and fills a $T^2$ or $T^3$.
Thus $Y$ cannot be a four-dimensional manifold.
For example,
$Y^{p,q}$ is quasi-regular only when there is an integer $n$ such that
$4p^2-3q^2=n^2$.  In these cases, the $\psi$ direction closes
with some periodicity $2\pi s/3$.
Then we obtain a good $b$-deformation by
orbifolding along the $\psi$ direction by $\IZ_s$,
although a small technical issue still remains, which  is discussed further
in appendix \ref{orbifolding}.

%In the case of $L^{p,q,r}$ ...
The main point we would like to raise is
the fact that although locally the $b$-deformation is allowed
the mere presentation of a Sasaki-Einstein manifold
in the form of (\ref{4cycle}) does not guarantee that the manifold
exists globally. In fact, in several cases one needs to find appropriate alternative coordinate in which to demonstrate
the existence of the SE manifold.
However, it is clear that if the base manifold exist for $b=0$ it also exists for $b\ne 0$.

The result can also be expressed in the following way.

As we saw, the blow-up four-cycles are governed by the internal
points on the diagram, where the point specifies which direction
of $T^3$ is degenerating along the four-cycle. The Reeb vector $R$
also defines a point on the toric diagram, and when the universal
deformation is possible, $R$ degenerates along the four-cycle.
Thus, $R$ must be on the internal integral points in order to do
the universal $b$-deformation. The orbit of the Reeb vector closes
or not according to whether $R$ lies on a rational point or an
irrational point. In the former case, we can take a suitable
orbifold to make the vector to sit on an internal integral point,
while in the latter there is no way to make it integral.

%%%%%%%%%%%%%%%%%%%%%%%%%%%%%%%%%%%%%%%%%%%%%%%%%%%%%%%%%%%%%%%%
\section{Fractional branes and $b$-deformation}\label{cascade}
%%%%%%%%%%%%%%%%%%%%%%%%%%%%%%%%%%%%%%%%%%%%%%%%%%%%%%%%%%%%%%%%%%%%

Given a quiver diagram describing a conformal field theory, a
simple way to upset the vanishing of the beta functions is by
changing the rank of some of the gauge groups. In the supergravity
side this operation corresponds to the introduction of fractional
branes. This is a universal mechanism discussed in
\cite{kn,ks,kt}. The addition of fractional branes generalizes to
many of the six-dimensional metrics discussed here. That is, the
ten-dimensional supergravity solution based on them accommodates
fractional D3 branes. In this section we show that {\it if a
Calabi-Yau cone admits fractional D3 branes then its deformation
by a parameter $b$ also admits fractional branes,} where
the fractional brane corresponds to D5 branes
wrapping a two-cycle in the blown-up four-cycle.

The presence of fractional branes manifest itself in the gravity
theory by the appearance of an imaginary self-dual 3-form flux
$G_3$. For the cone over Sasaki-Einstein metrics we have that $M$
fractional branes are described by \be \label{g3} G_3=M\left
(\frac{dr}{r}+ie^{\psi}\right )\w \o , \ee where $\o $ is an
appropriately chosen 2-form on the four  dimensional base. Suppose
such a construction of $G_3$ is allowed for a metric of the form
\be \label{met1} ds^2=dr^2+r^2(e^{\psi})^2+r^2ds_4^2 , \ee with
$ds_4^2$ a {\K}-Einstein metric. Then, the $b$-deformed metric \be
\label{met4} ds^2=\k (r)^{-1}dr^2+{\k}r^2(e^{\psi})^2+r^2ds_4^2,
\ee where $\k =1-b^6/r^6$, also admits an imaginary self-dual
3-form flux. In other words, the $b$-deformation does not
interfere with the property of the cone metric to admit both
regular and fractional branes.
\par
One can write an imaginary self-dual 3-form for the cone metric \eqref{met1},
if there exists an anti self dual $(1,1)$-form on the
four dimensional base satisfying \bea
d\o &=&0,\nn\\
J\w \o &=& 0,\nn\\
\label{p1} *_4\o &=& -\o, \label{cond}
 \eea
where $J$ is the {\K} form and $*_4$ is the Hodge dual taken with
respect to the four dimensional metric. Then the relevant 3-form
for $M$ fractional D3 branes is given precisely by \eqref{g3}.
The relevant properties of \eqref{g3} are: \bea
dG_3&=&0, \nn \\
\label{p2} {*}_6G_3 &=&iG_3 . \eea
They can be verified using
\eqref{p1} as: \bea dG_3&=& M[-r^{-1}dr\w d\o +ide^{\psi}\w \o
-ie^{\psi}\w d\o ]=0,\nn\\
{*}_6G_3&=&M[e^{\psi}\w *_4\o-ir^{-1}dr\w *_4 \o ]\nn\\
 &=& i M[r^{-1}dr\w \o +ie^{\psi}\w \o ]\nn\\
&=&iG_3,\nn\\
\eea where in the first line we used that $de^\psi=2J$. For the
$b$-deformed metric \eqref{met4}, we simply take \be G_3=M\left
(\frac{dr}{\k r}+ie^{\psi}\right )\w \o.\label{modG3}
 \ee
Then proceeding along exactly similar lines, one finds that this
3-form satisfies the properties \eqref{p2} for the
 metric \eqref{met4}. This construction has been used to obtain imaginary
self-dual 3-form for certain cases, see for example
\cite{pt,pal,sfetsos,ks,HEK,ben,lpqr,resolved}.

In a neighborhood of $r\sim b$, \eqref{modG3} can be
rewritten as \begin{equation}
G_3=M (\frac{du}u + ie^\psi)\wedge \omega,
\end{equation}where $u$ is a variable introduced in \eqref{udef}.
Take a two-cycle $C$ dual to $\omega$ on the blown-up four-cycle
so that $\int_C \omega=0$.
Let us take a small circle $c$ around the origin $u=0$
which winds around $\psi$ direction once. Then
\begin{equation}
\int_{c\times C}G_3=2\pi,
\end{equation}which means that the self-dual three-form field strength is
sourced by the D5 brane wrapped on the two-cycle dual to $C$ and
representing $\omega$.
Thus, the fractional D3 brane is now realized as
a D5 brane wrapping around the two-cycle.

%%%%%%%%%%%%%%%%%%%%%%%%%%%%%%%%%%%%%%%%%%%%%%%%%%%%%%%%%%%%%%%%%%%%%
\section{Conclusion}
%%%%%%%%%%%%%%%%%%%%%%%%%%%%%%%%%%%%%%%%%%%%%%%%%%%%%%%%%%%%%%%%%%%%%
In this paper we have studied a K\"ahler deformation of Calabi-Yau
spaces that can be obtained as cones over Sasaki-Einstein spaces.
This is a very large class of spaces that have been playing an
important role in the AdS/CFT.  We have described this deformation
explicitly in terms of toric geometry and studied its field theory
dual. We showed that this deformation corresponds to giving a
vacuum expectation value to a scalar dimension six operator with
vanishing $R$-charge and gave a suggestion for the the precise
form of this operator. Interestingly, a probe string in these
backgrounds signals area-law confinement (see appendix
\ref{confinement} for details). Thus, {\it we provide a universal
K\"ahler deformation that induces confinement in a large class of
quiver gauge theories}. We have also discussed some interesting
relations with a better-studied K\"ahler deformation called the
small resolution in the case of a $\mathbb{Z}_2$ orbifold of the
conifold, the so-called local $\IP^1 \times \IP^1$.

Some open questions remain. For example, we would like to improve
our understanding of the K\"ahler deformation from the algebraic
point of view. We did not find a fully satisfactory connection
between the  gauged linear sigma model analysis and the
supergravity prediction. Analogously, we would like to understand
better the role of the dimension six operators at the level of
field theory, possibly directly in the quiver diagram. We hope to
return to some of these questions in future work.

We argued that there are as many K\"ahler deformations
corresponding to blowing up 4-cycles as internal points in the
toric diagram. In this paper we have considered turning on only
one of them. It would be interesting to understand the general
case. Presumably, the general case corresponds to generalizations
of the metric of equation (\ref{PZT}) in the same way as the
Gibbons-Hawking metric generalizes the Eguchi-Hanson metric. What
is needed would be the Gibbons-Hawking metric with $U(1)\times
U(1)$ isometries, i.e. when all the point-charges lie in a line.
Similarly, it would be interested to consider the general case of
K\"ahler deformations corresponding to blowing up 2-cycles.

%%%%%%%%%%%%%%%%%%%%%%%%%%%%%%%%%%%%%%%%%%%%%%%%%%%%%%%%%%%%%%%%%%%%%%
\section*{Acknowledgments}
%%%%%%%%%%%%%%%%%%%%%%%%%%%%%%%%%%%%%%%%%%%%%%%%%%%%%%%%%%%%%%%%%%%%%%%%%%%%%%
We would like to thank D. Gaiotto, A. King, I. Klebanov, C. N\'u\~nez, T.
Okuda, C. R\"omelsberger, J. Sonnenschein, A. Tseytlin and J. Walcher for discussions. We
are particularly thankful to Ami Hanany
for many insightful comments. SB, LAPZ and YT  are grateful to the
organizers and participants of the KITP program ``Mathematical
Structures in String Theory'' for  hospitality and a very stimulating atmosphere.
This research was supported in part by Department of Energy under
grant DE-FG02-95ER40899 to the University of Michigan and the
National Science Foundation under Grant No. PHY99-07949 to the
Kavli Institute for Theoretical Physics.
YT is supported in part by the JSPS predoctoral fellowship.

%%%%%%%%%%%%%%%%%%%%%%%%%%%%%%%%%%%%%%%%%%%%%%%%%%%%%%%%%%%%%%%%%%%%%%
\appendix
%%%%%%%%%%%%%%%%%%%%%%%%%%%%%%%%%%%%%%%%%%%%%%%
%%%%%%%%%%%%%%%%%%%%%%%%%%%%%%%%%%%%%%%%%%%%%%%%%%%%%%%%%%%%%%%%%%%%%%%%%
\section{Ricci flatness}\label{ricci}
%%%%%%%%%%%%%%%%%%%%%%%%%%%%%%%%%%%%%%%%%%%%%%%%%%%%%%%%%%%%%%%%%%%%%%%%%
In this appendix, we explicitly show that the metric is Ricci
flat. This fact was established very generally in \cite{pp}, but
we will use some of the intermediate results in the main text. In
particular, arguments about supersymmetry rely on the explicit
form of the spin connection.
\be \label{metric}
 ds^2= A^2(r)dr^2+B^2(r)(d\psi+\s)^2+C^2(r)ds_4^2
 \ee
 The one forms are
 \bea
{e}^r&=&A(r)dr\nn\\
{e}^{\psi}&=&B(r)(d\psi+\s )\nn\\
{e}^i&=&C(r){\hat{e}}_i\;\;\;\;(i=1..4)
 \eea
They are chosen such that the metric is orthonormal. $d\s =2J$ and
$J=\frac{1}{2}J_{ij}\hat{e}^i\w \hat{e}^j$. Superscript hat
denotes the quantities with respect to four-dimensional base. The
one forms are
 \bea
 d{e}^r&=&0\nn\\
 d{e}^{\psi}&=&\frac{B'}{AB}{{e}^r}\w
 {{e}^{\psi}}+\frac{B}{C^2}J_{ij}e^i\w e^j\nn\\
d{e}^i&=&\frac{C'}{AC}{{e}^r}\w {{e}^i} -C^{-1}{\hw}{^i}_{jk}{e}^k
\w {e}^j
 \eea
 Prime means derivative with respect to $r$. ${\hw}{^i}_{j}={\hw}{^i}_{jk}\hat{e}^k
 $.
 For $de^a=-\frac{1}{2}C{_{bc}}^ae^b\w e^c$, we get
 \bea
 C{_{ab}}^r&=&0\nn\\
C{_{r\psi}}^{\psi}&=&-\frac{B'}{AB}\nn\\
C{_{ij}}^{\psi}&=&-\frac{2 B}{C^2}J_{ij}\nn\\
C{_{ri}}^i&=&-\frac{C'}{AC}\nn\\
C{_{jk}}^{i}&=&-\frac{1}{C}{\hw}{^i}_{jk} \eea
 Then the spin-coefficients are
\bea {\o}_{r\psi}&=&-\frac{B'}{AB}{e}^{\psi}\nn\\
{\o}_{ri}&=&-\frac{C'}{AC}{e}^{i}\nn\\
{\o}_{\psi i}&=&\frac{B}{C^2}J_{ij}e^j \nn\\
{\o}_{ij}&=&{\hw}_{ij} -\frac{B}{C^2}J_{ij}{e}^{\psi}
 \eea
 This leads to Ricci two form components using
 $
 {\t}_{ij}=dw_{ij}+w_{ik}\w w{^{k}}_{j}
 $
The indices are raised and lowered in this case by
${\delta}{^{i}}_{j}$.
 \bea
{\t}_{r\psi}&=&-\frac{1}{AB}\left (\frac{B'}{A}\right
)'{{e}^r}\w{{e}^{\psi}} -\frac{1}{AC^3} ( B'C-BC')J_{ij}e^i\w e^j\nn\\
{\t}_{ri}&=&-\frac{ 1}{AC}\left (\frac{C'}{A}\right
)'{{e}^r}\w{{e}^{i}} +\frac{1}{AC^3}(B'C-BC')J_{ij}{e}^{j}\w{e}^{\psi}\nn\\
{\t}_{\psi i}&=& \frac{1}{AC^3}(B'C-BC')J_{ij}e^r\w e^j +\left
[\left (\frac{B}{C^2}\right )^2-\frac{B'C'}{A^2BC}\right
]e^{\psi}\w
e^i\nn\\
&&+\frac{B}{C^2}(DJ_{ij})\w
e^j\nn\\
{\t}_{ij}&=&{\hat{\t}}_{ij}-\frac{2}{AC^3}(B'C-BC')J_{ij}e^r\w
e^{\psi}\nn\\
&&-\left (\frac{B}{C^2}\right )^2(J_{ij}J_{kl}+J_{ik}J_{jl})e^k\w
e^l -\left (\frac{C'}{AC}\right )^2e^i\w e^j\nn\\
&&-\frac{B}{C^2}(DJ_{ij})\w e^{\psi}\nn\\
\tr{where }DJ_{ij}&=&dJ_{ij}-J_{ik}{\hw}{^k}_j+{\hw}_{ik}J{^k}_j
 \eea
Let us invoke here $DJ_{ij}=0$. Since
${\t}{^a}_b=\frac{1}{2}R{^a}_{bcd}e^c\w e^d$, one can get
 \bea
 R_{r\psi r\psi}=-\frac{1}{AB}\left (\frac{B'}{A}\right )'&&
 R_{r\psi ij}=-\frac{2}{AC^3}(B'C-BC')J_{ij}\nn\\
 R_{riri}=-\frac{1}{AC}\left (\frac{C'}{A}\right )'
&&R_{r ij\psi}=\frac{1}{AC^3}(B'C-BC')J_{ij}\nn\\
R_{\psi irj}=\frac{1}{AC^3}(B'C-BC')J_{ij}&& R_{\psi
ii\psi}=\frac{B'C'}{A^2BC}-\left (\frac{B}{C^2}\right )^2\nn\\
R_{ijkl}=C^{-2}{\hat{R}}_{ijkl}-\left (\frac{B}{C^2}\right
)^2(2J_{ij}J_{kl}+J_{ik}J_{jl}-J_{il}J_{jk})&-&\left
(\frac{C'}{AC}\right
)^2({\d}_{ik}{\d}_{jl}-{\d}_{il}{\d}_{jk})\nn\\
R_{ijr\psi}=-\frac{2}{AC^3}(B'C-BC')J_{ij}&&
 \eea
 This gives
 \bea
 R_{rr}&=&-\frac{1}{AB}\left (\frac{B'}{A}\right )'-
 \frac{4}{AC}\left (\frac{C'}{A}\right
 )'\nn\\
 R_{\psi\psi}&=&-\frac{1}{AB}\left (\frac{B'}{A}\right )'
 -\frac{4B'C'}{A^2BC}+4\left (\frac{B}{C^2}\right )^2\nn\\
R_{ii}({\textrm{ no sum }})&=&-\frac{1}{AC}\left
(\frac{C'}{A}\right
 )'-\frac{B'C'}{A^2BC}+\left (\frac{B}{C^2}\right
 )^2+C^{-2}{\hat{R}}_{ii}+2\left (\frac{B}{C^2}\right
)^2{\d}_{ii}-3\left (\frac{C'}{AC}\right )^2{\d}_{ii}\nn\\
R_{ij}\;\;(i\neq j)&=&C^{-2}{\hat{R}}_{ij}
 \eea
 For Einstein base, ${\hat{R}}_{ij}$ is proportional to metric.
 Since we are doing calculations in orthonormal frame, it vanishes
 for $i\neq j$. The Ricci flatness conditions are
 \bea
-\frac{1}{4AB}\left (\frac{B'}{A}\right )'-
 \frac{1}{AC}\left (\frac{C'}{A}\right
 )'=0\nn\\
-\frac{1}{4AB}\left (\frac{B'}{A}\right )'
 -\frac{B'C'}{A^2BC}+\left (\frac{B}{C^2}\right )^2=0\nn\\
\label{eqn}
 C^{-2}{\l}+\frac{1}{2AB}\left (\frac{B'}{A}\right )'
+2\left (\frac{B}{C^2}\right )^2-3\left (\frac{C'}{AC}\right
)^2&=&0
 \eea
 where $\hat{R}_{ij}={\l}\hat{g}_{ij}$ for the Einstein base.
 A general solution for these equations was obtained in \cite{pp}.
 The equations are easier to handle if one chooses $AB=c$ (a
 constant) in the beginning. The solution is
\be \label{met2}
d\hat{s}^2=(1-r^2)^2P(r)^{-1}dr^2+c^2(1-r^2)^{-2}P(r)(d\psi
+\s)^2+c(1-r^2)ds_{4}^2 \ee where, $P(r)$ is a polynomial in $r$.
If one imposes the condition for the metric to be {\K}, then it
takes the form \bea d\hat{s}^2&=&
U^{-1}d{\rho}^2+U{\rho}^2(d\tau +\s)^2+{\rho}^2ds_4^2\nn\\
U&=&\frac{\l}{6}+c_0{\rho}^{-6}
 \eea
 $c_0$ is a constant.
 If now one sets $\l =2(n+1)$, then
\be
 d\hat{s}^2=\left [1-\left (\frac{b}{\rho}\right )^{6}\right
 ]^{-1}d{\rho}^2+\left [1-\left (\frac{b}{\rho}\right )^{6}\right
 ]{\rho}^2(d\psi +\s)^2+{\rho}^2ds_4^2
\ee So the b-deformed metric is the most general Calabi-Yau metric
of type \ref{metric}.
%%%%%%%%%%%%%%%%%%%%%%%%%%%%%%%%%%%%%%%%%%%%%%%%%%%%%%%%%%%%%%%%%%%%%%%%%%%%%%
\section{Supersymmetry}\label{susy}
%%%%%%%%%%%%%%%%%%%%%%%%%%%%%%%%%%%%%%%%%%%%%%%%%%%%%%%%%%%%%%%%%%%%%%%%%%%%%%%%%%
 We will write down the integrability conditions for these metrics
 to be supersymmetric. We check here that deformation by
 b-parameter do not spoil supersymmetry.
 The integrability conditions $R_{abcd}{\G}^{cd}\v
 =0$ written explicitly are
 \bea
 -\frac{1}{AB}\left (\frac{B'}{A}\right
 )'{\G}^{r\psi}\v -\frac{2}{AC^3}(B'C-BC')J_{ij}{\G}^{ij}\v =0\nn\\
- \frac{1}{AC}\left (\frac{C'}{A}\right
 )'{\G}^{ri}\v +\frac{1}{AC^3}(B'C-BC')J{^i}_{j}{\G}^{j\psi}\v =0\nn\\
\frac{1}{AC^3}(B'C-BC')J_{ij}{\G}^{rj}\v +\left
(\frac{B'C'}{A^2BC}-\frac{B^2}{C^4}\right ){\G}^{i\psi}\v =0\nn\\
C^{-2}\hat{R}_{ijkl}{\G}^{kl}\v-2\frac{B^2}{C^4}(J_{ij}J_{kl}+J_{ik}J_{jl}){\G}^{kl}\v-2\left
(\frac{C'}{AC}\right
)^2{\G}^{ij}\v-\frac{2}{AC^3}(B'C-BC')J_{ij}{\G}^{r\psi}\v=0
\nn\\\eea For the case of $A=1, B=C=r$, the only nontrivial
equation is \be \label{e1}
\hat{R}_{ijkl}{\G}^{kl}\v-2(J_{ij}J_{kl}+J_{ik}J_{jl}){\G}^{kl}\v-2\G
_{ij}\v =0 \ee For $A={\k} ^{-1/2}, B= \k ^{1/2}r, C=r$ with $\k
=1-b^6/r^6$, they are
\bea\label{e2} [4\G_{r\psi}-J_{ij}\G ^{ij}]\v=0\\
\label{e3}
\G ^{ri}\v-J{^i}_j\G ^{j\psi}\v =0\\
\label{e4}
\G ^{i\psi}\v+J_{ij}\G ^{rj}\v =0\\
\hat{R}_{ijkl}\G ^{kl}\v -2\k (J_{ij}J_{kl}+J_{ik}J_{jl})\G
^{kl}\v -2\k \G ^{ij}\v-2r{\k}'J_{ij}\G ^{r\psi}\v=0
 \eea
 The last equation can be simplified further if we assume equation
 \ref{e1},
 \be
 \label{e5}
 (J_{ij}J_{kl}+J_{ik}J_{jl})\G ^{kl}\v+\G _{ij}\v-6J_{ij}\G ^{r\psi}\v=0
 \ee
 The equation (\ref{e3}) gives us the projections. Using it, equations
 \ref{e2}, \ref{e4} and \ref{e5} can then be satisfied. So, we
 conclude that the b-deformation do not spoil the supersymmetry.

%%%%%%%%%%%%%%%%%%%%%%%%%%%%%%%%%%%%%%%%%%%%%%%%%%%%%%%%%%%%%%%%%%%%%%%%%%%%%%
\section{Compactification}\label{action}
%%%%%%%%%%%%%%%%%%%%%%%%%%%%%%%%%%%%%%%%%%%%%%%%%%%%%%%%%%%%%%%%%%%%%%%%%%%%%%%%%%
In this section, we look at the metric perturbations same as in
\cite{kt}. We will try to find their masses for the general case
of b-deformed metrics. \bea \label{met11}
ds_{10}^2&=&L^2[M^2{ds_5}^2+ds_{5'}^2]\\
ds_5^2&=&du^2+e^{2A}dx_{\mu}dx^{\mu}\nn\\
ds_{5'}^2&=&e^{2B}(d\psi +\sigma)^2+e^{2C}(e_a^2)\nn \eea where
$M=-\frac{1}{3}(B+4C)$ and $A, B$ and $C$ are arbitrary functions
of $u=\ln r$. L is a constant. $a$ runs from 1 to 4 and $e^a$
together form a 4 dimensional {\K}-Einstein metric with {\K}- form
$J$. $\s$ is such that $d\s =2J$. Latin indices $a,b,c,..$ runs
over this 4 dimensional base, while Greek indices $\mu ,\nu, ...$
denote indices over the Poincare metric $dx_{\mu}dx^{\mu}$. The
type IIB solution is accompanied with a five form self dual flux.
\be F_5=\cf +*\cf\;\;\;\cf=Q(d\psi +\s)\w e_1\w e_2\w e_3\w e_4\ee
Q is a constant. The Riemann components are
\bea R_{\mu\nu\mu\nu}&=&-\frac{1}{L^2M^4}(MA'+M')^2\nn\\
R_{\mu u\mu
u}&=&-\frac{1}{L^2M^4}(-{M'}^2+MM'A'+M^2A''+MM''+M^2{A'}
^2)\nn\\
R_{\mu\psi \mu\psi}&=&-\frac{B'}{L^2M^3}(MA'+M')\nn\\
R_{\mu a\mu a}&=&-\frac{C'}{L^2M^3}(MA'+M')\nn\\
\eea Rest of the Riemann tensor components are similar to the
corresponding ones in appendix A.
 Superscript prime here means derivative with respect to $u$.
Then the Ricci scalar for the metric \ref{met11} is \bea
R&=&\frac{1}{L^2e^{2C}}\hat{R}-\frac{1}{L^2M^4}(4M^2\{2A''+5{A'}^2\}+{M'}^2+2M'A'+2MM''+{B'}^2+4{C'}^2)\nn\\&&-\frac{4}{L^2}e^{2B-4C}
\eea The determinant of the metric in terms of one forms used in
\ref{met11} is \be g_{10}=L^{20}M^4g_5\;\;\;\;g_5=e^{8A} \ee $g_5$
is the determinant of $ds_5^2$. Using variables $q$ and $f$
defined as $B+4C=\frac{15}{2}q$ and $B-C=-5f$, the action \be
S=-\frac{1}{2{\k}^2}\int\left [ d^{10}x
\sqrt{g_{10}}R-\frac{1}{2}\cf\w *\cf\right ]\ee can be written as
\bea S&=& -\frac{1}{2k_5^2}\int d^5 x \sqrt{g_5}\left
[R_5+M^2e^{-2C}\hat{R}-(30{q'}^2+20{f'}^2)-4M^2e^{2B-4C}\right
]\nn\\&&-\frac{L^8}{\k ^2}\int d^{5}x \left
(\frac{M'}{M}e^{4A}\right )'+\frac{1}{4\k ^2}\int\cf\w *\cf \eea
The penultimate term being a total derivative can be removed from
consideration. The action becomes \be S=-\frac{2}{\k _5^2}\int
d^{5}x \sqrt{g_5}\left
[\frac{1}{4}\hat{R}-\frac{1}{2}(15{q'}^2+10{f'}^2)-e^{-8q}(e^{-12f}-\l
e^{-2f})-\frac{1}{8}Q^2e^{-20q}\right ] \ee where $\l$ is defined
for the 4-dimensional base as $\hat{R}_{ab}=\l g_{ab}$. For
Sasaki-Einstein metrics, $\l$ is 6. $Q$ can be fixed to be 4 so
that AdS space has unit radius (For L=1). Then for small enough
$q$ and $f$, the lowest order terms in the potential are \bea
V&=&e^{-8q}(e^{-12f}-\l
e^{-2f})+\frac{1}{8}Q^2e^{-20q}\nn\\&=&(3-\l)+(\l
-6)[-16qf+2f+8q-32q^2-2f^2]+60f^2+240q^2+... \nn\eea The action
for the case $\l =6$ becomes \be S=-\frac{2}{\k _5^2}\int d^{5}x
\sqrt{g_5}\left
[\frac{1}{4}\hat{R}-\frac{1}{2}\{15({q'}^2+32q^2)+10({f'}^2+12f^2)\}\right
 ]+\tr{constant+ higher order terms}
\ee So the perturbations $q$ and $f$ have masses given by $m^2_q=
32$ and $m^2_f=12$ respectively, when $\l =6$. For $\l \ne 6$, these
are not the correct perturbations to be considered.

%%%%%%%%%%%%%%%%%%%%%%%%%%%%%%%%%%%%%%%%%%%%%%%%%
\section{Orbifold singularities on $b$-deformation}\label{orbifolding}
%%%%%%%%%%%%%%%%%%%%%%%%%%%%%%%%%%%%%%%%%%%%%%%%%
In this appendix we carry out a preliminary study of the orbifold
singularities on the blown-up four-cycle in the $b$-deformation in
the quasi-regular case.

Let $X$ be a quasi-regular Sasaki-Einstein space and let $U(1)$ to
act on $X$ by the shift along the Reeb vector. The metric on $X$
can locally be canonically written in the form
\begin{equation}
ds_X^2=(d\psi +\sigma)^2+ds_B^2
\end{equation}
where $\partial_\psi$ is the Reeb vector. Suppose $\psi$ has
periodicity $2\pi k/3$ on the generic points. As discussed in
section \ref{general}, to have a good $b$-deformation we need to
have the periodicity $\psi\sim\psi+2\pi/3$. Thus, we need to
orbifold by $\IZ_k$. For a regular Sasaki-Einstein this is all we
need, but for a quasi-regular Sasaki-Einstein, we need to worry
whether or not this orbifolding preserves the covariantly constant
spinor.

Consider a point $p$ on $X$ where the Reeb vector closes with
period $2\pi k/(3n)$. Since $X$ is smooth, we can take, near $p$,
a coordinate patch parametrized by $\psi$ and two complex
variables $\vec z=(z_1,z_2)$ with almost flat metric so that $p$
is at $z_1=z_2=\psi=0$. The $2\pi k/(3n)$ periodicity at $p$ means
that there is an element $g\in U(2)$ with $g^n=1$ such that $X$
near $p$ is given by the identification \begin{equation} (\vec
z,\psi)\sim (g\vec z,\psi+\frac{2\pi k}{3n}).
\end{equation} Orbifolding by $\IZ_k$ changes this to\begin{equation}
(\vec z,\psi)\sim (g^{1/k}\vec
z,\psi+\frac{2\pi}{3n}).\label{identif}
\end{equation}
Adding the $u$ direction combines with the direction $\psi$ to
make another complex coordinate $z_3=ue^{i3\psi}$ so that now the
metric is $|dz_1|^2+|dz_2|^2+|dz_3|^2$. Then the identification
(\ref{identif}) is given by \begin{equation} (\vec z,z_3)\sim
(g^{1/k}\vec z,e^{i2\pi/n}z_3).
\end{equation}
Thus we need to have $e^{2\pi i/n}\mathrm{det}\ g^{1/k} =1$ to
preserve the Calabi-Yau condition.

Let us check in the case of the first quasi-regular $Y^{p,q}$,
which is $p=7$, $q=3$. We use the notation of \cite{se}. Then
$\ell=3/20$, the Reeb vector is \begin{equation}
\partial_{\psi',theirs}=\partial_{\psi,theirs}- \frac{10}9 \partial_{\alpha/\ell},
\end{equation}and $\psi$, $\alpha/\ell$ both have periodicity $2\pi$.
Beware that $\psi'_{theirs}=3\psi_{ours}$.

For the orbifold point at $y=y_1$ and $\theta=0$, the variables
$z_i$ above are given in the variables in \cite{se} by
\begin{equation}
z_1=\theta e^{i\phi},\quad z_2=R e^{i(\psi-\phi)},\quad z_3=u e^{i
\frac9{35}(\frac{\alpha}{\ell} +5 (\psi-\phi))}
\end{equation}
Thus $n=35$, $k=9$ and $g=\mathrm{diag}(1,e^{2\pi i 9/35})$. so
that the orbifolding acts by \begin{equation} (z_1,z_2,z_3)\sim
(z_1,e^{2\pi i/35}z_2,e^{2\pi i/35}z_3).
\end{equation}

Thus we have an ALE singularity in the blownup geometry.
The same analysis can be carried out on the other three vertices
and for other quasi-regular $Y^{p,q}$s.

 %%%%%%%%%%%%%%%%%%%%%%%%%%%%%%%%%%%%%%%%%%%%%%
\section{Confinement}\label{confinement}
%%%%%%%%%%%%%%%%%%%%%%%%%%%%%%%%%%%%%%%%%%%%%%%%%
In this section, we will go through the steps which shows
confinement behavior for strings in b-deformed background. The
calculations were done in \cite{pt} and repeated here for sake of
completeness.
The Nambu-Goto string action for the Wilson loop is \be S=\int
d\tau d\sigma\sqrt{-\tr{det}g_{ab}}\ee where $g_{ab}$ is the
embedded metric on the string worldsheet, and  $a,b=1,2$.
We denote the
background metric by $G_{MN}$ ,$M,N=1\ldots10$.
We assume that string
extends only in radial direction. In static gauge, i.e. $\tau
=x_0$ and $x_1 = \sigma$, the worldsheet metric becomes \be
g_{ab}=\left [\begin{array}{cc}G_{00}&0\\0&G_{11}+G_{rr}(d _x
r)^2\end{array}\right ]\ee This leads to the action \be S=\int
d\tau\int dx\sqrt{h^{-1}+\k ^{-1}(d_xr)^2}=T\int
dr\sqrt{(d_rx)^2h^{-1}+\k ^{-1}}\ee  Since the Lagrangian is
independent of variable $x$, one has a constant of motion
\be\frac{\p L}{\p (d_rx)}=\frac{h^{-1}}{\sqrt{h^{-1}+\k
^{-1}(d_xr)^2}}=c_0 \;\;(\tr{say}) \ee Rearranging, one gets \be
dx=\frac{c_0hdr}{\sqrt{\k(1-c_0^2h)}} \ee and can calculate the
length of Wilson loop to be \be\frac{l}{2}=\int dx
=\frac{L^2}{\sqrt{2}b}\int
^{\infty}_{y_{*}}dy\frac{y}{\sqrt{y^3-1}}\frac{f(y)}{\sqrt{f(y_*)-f(y)}}\ee
Here, the action is re-expressed in terms of variable $y=r^2/b^2$.
The constant factor in front of warp factor \ref{warp} as
$h=\frac{2L^4}{b^4}f(y)$. $y_*$ is the point where $c_0^2h=1$ or,
$h\k ^{-1}(\p _xr)^2$ vanishes. This is the turning point of the
string. The string does not explore regions lying further interior
in the bulk. At this point, $f(y_*)=\frac{b^4}{2L^4c_0^2}$. The
energy of the string is \be
E=\frac{S}{T}=\frac{b^3}{2^{3/2}L^2c_0}\int
_{y_*}^{\infty}\frac{ydy}{\sqrt{y^3
-1}}\frac{1}{\sqrt{f(y_*)-f(y)}}\ee One evaluates these integrals
under the assumption that large contributions come from region
close to $y=y_*$. In such a case, $f(y)$ is approximated as
$f(y)=f(y_*)+f'(y_{*})(y-y_*)$. This leads to \bea
\frac{l}{2}&=&\frac{L^2}{\sqrt{2}b}\frac{f(y_*)}{\sqrt{f'(y_*)}}\int\frac{ydy}{\sqrt{(y^3-1)(y-y_*)}}\nn\\
E&=&\frac{b^3}{2^{3/2}L^2c_0}\frac{1}{f'(y_*)}\int\frac{ydy}{\sqrt{(y^3-1)(y-y_*)}}\nn\eea
So, one finds the relation between energy and length of the Wilson
loop to be \be\frac{2E}{l}=\frac{b^4}{2L^4c_0f(y_*)}\Rightarrow
E\sim \frac{c_0}{2}l\ee This shows the confining area law behavior
for the spatial Wilson loop.

%%%%%%%%%%%%%%%%%%%%%%%%%%%%%%%%%%%%%%%%%%%%%%%%%%%%%%%%%%%%%%%%%%%%%%%%%
\section{Probe D7 branes: Polarization and resolution of singularities}\label{d7}
%%%%%%%%%%%%%%%%%%%%%%%%%%%%%%%%%%%%%%%%%%%%%%%%%%%%%%%%%%%%%%%%%%%%%%%%%
The solution has a singularity in the infrared region. In the
context of the AdS/CFT, there are  many cases where singularities
in the infrared turn out to be resolved or at least understood in
terms of brane sources. We  give evidence supporting the claim
that the singularity does not affect the studies of various
properties of the gauge theories. A similar behavior has been
amply shown in the case of the Constable-Myers background
\cite{cm}. In particular, \cite{evans} show that various probes
never reach the singularity and more interestingly, various
properties of the IR regime are well defined including the pattern
of chiral symmetry breaking, quark condensate and others. These
properties  turned out to be independent of the existence of a
singularity in the infrared. We leave a detailed analysis of the
properties of this class of solution to the future but the
solution is confining according to the arguments in previous
section.

\begin{figure}[h!]
\begin{center}
\epsfig{file=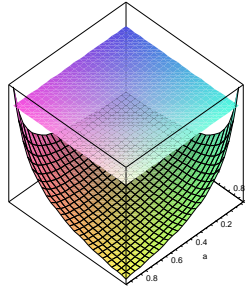,width=4cm} \caption{$r_0^6/b^6$ is plotted along
height of above cube against $a$ and $c$. The flat surface is
where $r_0=b$. Here , we take $\frac{L^4e^{\Phi}}{36}=1$.
\label{plot}}
\end{center}
\end{figure}

The background presented in section \ref{d3branes} has a
singularity in the region $r\to b$ where \be h(r\to b)
=-\frac{2L^4}{3b^4}\ln \left(\frac{r}{b}-1\right) +\ldots, \ee It
was noted in the case of the conifold \cite{pt} that this
singularity is a curvature singularity. The logarithmic behavior
suggests the presence of a 7-brane. Note also that the logarithmic
behavior suggest a space of codimension two. With this intuition,
we turn to the question of the behavior of probe D7 branes in this
background. The general calculation in an arbitrary
Sasaki-Einstein space could be performed under some mild
assumptions. We will explicitly consider the case of the local
$\IP^1 \times \IP^1$ discussed in section \ref{kaehler}, but we
expect the results to be general. The basic setup is as follows:
we consider a supersymmetric D7 brane in the background with $b=0$
and concentrate on the effects caused by introducing a small
$b$-deformation. The supergravity background is approximately of
the form:
\begin{eqnarray}
ds^2&=&h^{-1/2}{\eta}_{\mu\nu}dx^{\mu}dx^{\nu}+h^{1/2}(dr^2+r^2(e^{\psi})^2+r^2ds_4^2),\nn\\
ds_4^2&=&(d\t _1^2+\sin ^2\t _1d\phi _1^2 )+(d\t _2^2+\sin ^2\t
_2d\phi _2^2 ),\nn\\
h&=&\frac{L^4}{r^4}\left (1+f(r)\right
)\;\;f(r)=\frac{2b^6}{5r^6}+\frac{b^{12}}{4r^{12}}+...\nn\\
 F_5&=&dC_4=d\chi _4+*d\chi _4, \qquad \chi
_4=\frac{1}{h}dx^0\w
dx^1\w dx^2\w dx^3\nn\\
 \end{eqnarray}
and a constant complex scalar $\tau =\frac{\t}{2\pi}+\frac{i}{g}$.
We assume other fields to be zero. The D7 probe brane in this
background extends parallel to the D3 brane and wraps $ds_4^2$.
The general action is \be S=\int d^8\xi
e^{-\Phi}[-\tr{det}(e^{\Phi/2}G_{ij}+F_{ij})]^{1/2}+\int
C_8+\frac{1}{2}\int F\w F\w C_4,
 \ee
 where $C_i$ fields are R-R forms. $G_{ij}$ is the pull-back of the ten-dimensional metric to the
D7 brane worldvolume and  $F_{ij}$ is the field strength
 of  a $U(1)$ gauge field on the worldvolume of the brane. A particular supersymmetric probe in this background is given by
the embedding  \cite{acr}:
\be
\label{embedding}
r^{-3}=r_0^{-3}\sin \t _1\sin \t _2, \qquad \psi =0.
\ee
We assume static gauge
for rest of the coordinates. Then the pull back
metric is
 \bea
 g_{ab}d{\xi}^{a}d{\xi}^{b}&=&h^{-1/2}dx_{\mu}dx^{\mu}+h^{1/2}r^2{ds'}^2_4,\\
{ds'}_4^2&=&\left (\frac{1}{9}\cot ^2\t
 _1+\frac{1}{6}\right )d\t _1^2+ \left (\frac{1}{9}\cos ^2\t
 _1+\frac{1}{6}\sin ^2\t _1\right )d\phi _1^2\nn\\
&&+\left (\frac{1}{9}\cot ^2\t
 _2+\frac{1}{6}\right )d\t _2^2+ \left (\frac{1}{9}\cos ^2\t
 _2+\frac{1}{6}\sin ^2\t _2\right )d\phi _2^2\nn\\
&& +\frac{2}{9}\cot \t_1\cot \t _2d\t _1d\t _2+\frac{2}{9}\cos \t _1\cos \t _2d\phi _1d\phi _2,
 \eea
 along with a non-trivial world volume gauge field $F_2$
 \be
  \label{F2} F_2= 2a\sin\t _1 d\t _1\w
d\phi _1+2c\sin \t _2d\t _2\w d\phi _2
 \ee
where $a$ and $c$ are arbitrary constants and $C_8=0$. After some
algebra, the action can be written as \be S=\int d^8\xi
\frac{r^4}{L^4(1+f(r))}[\sqrt{g_4}+4ac\sin \t _1\sin \t_2],
 \ee
where
$g_4=L^8e^{2\Phi}(1+f(r))^2g_1+L^4e^{\Phi}(1+f(r))g_2+a^2c^2\sin
^2\t _1\sin ^2\t _2$. Here, $g_1$ and $g_2$ are \bea
g_1&=&\frac{1}{2916\sin ^2\t _1\sin ^2\t _2}[\cos ^4\t_1\sin ^4\t
_2+2\cos ^2\t _1\sin ^2\t _1\cos ^2\t _2\sin ^2\t _2+3\cos ^2\t
_1\sin ^2\t _1\sin ^4\t _2\nn\\&&+\sin ^4\t _1\cos ^4\t _2+3\sin
^4\t _1\cos ^2\t _2\sin ^2\t _2+\frac{9}{4}\sin ^4\t
_1\sin ^4\t _2]\nn\\
g_2&=&-\frac{1}{81\sin ^2\t _1\sin ^2\t _2}[c^2\cos ^4\t _1\sin
^4\t _2 +a^2\sin ^4 \t _1\cos ^4\t _2+2ac\sin ^2\t _1\cos ^2\t
_1\sin ^2\t _2\cos ^2\t _2\nn\\&&+3(c^2\cos ^2\t _1\sin ^2\t
_1\sin ^4\t _2+a^2\sin ^4\t _1\cos ^2\t _2\sin ^2\t _2)
+\frac{9}{4}(c^2+a^2)\sin ^4\t _1\sin ^4\t _2]\eea
 Recall that in the
supersymmetric radial embedding  (\ref{embedding}) the minimal
radial distance of the D7 from the origin is given by $r_0$. This
action, as a function of $r_0$,  has minima at a non-trivial value
of $r_0$ given by solution of
 \be
\frac{4}{r}=f'\left
[\frac{1}{1+f(r)}-\frac{2L^8e^{2\Phi}(1+f(r))g_1+L^4e^{\Phi}g_2}{2\{g_4+4ac\sqrt{g_4}\sin
\t _1\sin \t _2\}}\right ]
 \ee
 If $b<<r$, then $f(r)\approx \frac{2b^6}{5r^6}$ and we can neglect $f(r)$ with respect to one.
 The equation has a solution for $r_0$ at
 \be
 b^6/r_0^6=\frac{10}{3}\frac{1}{\sin ^2\t _1\sin ^2\t _2}\left[\frac{g_4+4ac\sqrt{g_4}\sin \t _1\sin \t _2}
 {L^4e^{\Phi}(-g_2)-2a^2c^2\sin ^2\t _1\sin ^2\t _2-8ac\sqrt{g_4}\sin \t _1\sin \t _2}\right
 ]
 \ee
To arrive at a simple expression, we  further evaluate it for the
case when $\t _1=\t _2=\pi /2$,
 then $g_1=1/1296, g_2=-\frac{1}{36}(a^2+c^2)$, $g_4
 =\left (\frac{L^4e^{\Phi}}{36}-a^2\right )\left (\frac{L^4e^{\Phi}}{36}-c^2\right )$. The
 above expression becomes
 \be
 b^6/r_0^6=\frac{10}{3}\left [\frac{g_4+4ac\sqrt{g_4}}{\frac{L^4e^{\Phi}}{36}(a^2+c^2)-2a^2c^2-8ac\sqrt{g_4}}\right ]
 \ee
The above expression is independent of $L^4e^{\Phi}$, which can be
seen by a rescaling of $a$ and $c$. If $a^2\approx
\frac{L^4e^{\Phi}}{36}$ and $c\approx 0$ or vice-versa, then
$b^6/r_0^6 <1$. If $a,c$ both are chosen such that $a^2\approx
c^2\approx 0$,
 %\bea
 %\tr{Numerator}&\rightarrow& \frac{L^8e^{2\Phi}}{36^2}\nn\\
%\tr{Denominator}&\rightarrow& 0\nn
% \eea
 then $b^6/r_0^6>1$. So the ratio $b^6/r_0^6$ depends on the values $a$ and
 $c$. The quantity $r_0^6/b^6$ is plotted for various values of a
 and c in the plot \ref{plot}.
 \par

 The case $b^6/r_0^6 <1$ seems to be interesting. It hints that D7
probe brane can rest at a non-trivial value of $r_0>b$, ahead of
the
 problematic point of $r_0=b$. Since the probe brane do not venture
 through the strongly coupled region, predictions based on
 supergravity calculations should be reliable. For the configurations with these
 particular flux values, the open strings emanating from N D3
 branes do have a place to end in the bulk.


\begin{thebibliography}{99}
\bibitem{malreview}
%\cite{Maldacena:1997re}
%\bibitem{Maldacena:1997re}
  J.~M.~Maldacena,
 ``The large N limit of superconformal field theories and supergravity,''
  Adv.\ Theor.\ Math.\ Phys.\  {\bf 2} (1998) 231
  [Int.\ J.\ Theor.\ Phys.\  {\bf 38} (1999) 1113]
  [arXiv:hep-th/9711200].\\
  %%CITATION = HEP-TH 9711200;%%
%\cite{Aharony:1999ti}
%\bibitem{Aharony:1999ti}
  O.~Aharony, S.~S.~Gubser, J.~M.~Maldacena, H.~Ooguri and Y.~Oz,
 ``Large N field theories, string theory and gravity,''
  Phys.\ Rept.\  {\bf 323} (2000) 183
  [arXiv:hep-th/9905111].
  %%CITATION = HEP-TH 9905111;%%

\bibitem{wittengkp}
%\cite{Witten:1998qj}
%\bibitem{Witten:1998qj}
  E.~Witten,
``Anti-de Sitter space and holography,''
  Adv.\ Theor.\ Math.\ Phys.\  {\bf 2} (1998) 253
  [arXiv:hep-th/9802150].\\
  %%CITATION = HEP-TH 9802150;%%
%\cite{Gubser:1998bc}
%\bibitem{Gubser:1998bc}
  S.~S.~Gubser, I.~R.~Klebanov and A.~M.~Polyakov,
``Gauge theory correlators from non-critical string theory,''
  Phys.\ Lett.\ B {\bf 428} (1998) 105
  [arXiv:hep-th/9802109].
  %%CITATION = HEP-TH 9802109;%%

\bibitem{kw}
%\cite{Klebanov:1998hh}
%\bibitem{Klebanov:1998hh}
  I.~R.~Klebanov and E.~Witten,
``Superconformal field theory on threebranes at a Calabi-Yau
singularity,''
  Nucl.\ Phys.\ B {\bf 536} (1998) 199
  [arXiv:hep-th/9807080].
  %%CITATION = HEP-TH 9807080;%%


\bibitem{se}
%\cite{Gauntlett:2004zh}
%\bibitem{Gauntlett:2004zh}
  J.~P.~Gauntlett, D.~Martelli, J.~Sparks and D.~Waldram,
``Supersymmetric $\mathrm{AdS}^5$ solutions of M-theory,''
  Class.\ Quant.\ Grav.\  {\bf 21} (2004) 4335
  [arXiv:hep-th/0402153].\\
  %%CITATION = HEP-TH 0402153;%%
%\cite{Gauntlett:2004yd}
%\bibitem{Gauntlett:2004yd}
  J.~P.~Gauntlett, D.~Martelli, J.~Sparks and D.~Waldram,
 ``Sasaki-Einstein metrics on $S^2\times S^3$ ,''
  Adv.\ Theor.\ Math.\ Phys.\  {\bf 8} (2004) 711
  [arXiv:hep-th/0403002].
  %%CITATION = HEP-TH 0403002;%%

%\cite{Benvenuti:2004dy}
\bibitem{Benvenuti:2004dy}
  S.~Benvenuti, S.~Franco, A.~Hanany, D.~Martelli and J.~Sparks,
``An infinite family of superconformal quiver gauge theories with
Sasaki-Einstein duals,''
  JHEP {\bf 0506}, 064 (2005)
  [arXiv:hep-th/0411264].
  %%CITATION = HEP-TH 0411264;%%

\bibitem{cvetic}
%\cite{Cvetic:2005ft}
%\bibitem{Cvetic:2005ft}
  M.~Cveti\vv c, H.~Lu, D.~N.~Page and C.~N.~Pope,
 ``New Einstein-Sasaki spaces in five and higher dimensions,''
  Phys.\ Rev.\ Lett.\  {\bf 95} (2005) 071101
  [arXiv:hep-th/0504225].
  %%CITATION = HEP-TH 0504225;%%

\bibitem{lpqr}
%\cite{Martelli:2005wy}
%\bibitem{Martelli:2005wy}
  D.~Martelli and J.~Sparks,
``Toric Sasaki-Einstein metrics on $S^2 \times S^3$,''
  Phys.\ Lett.\ B {\bf 621} (2005) 208
  [arXiv:hep-th/0505027].
  %%CITATION = HEP-TH 0505027;%%

%\cite{Benvenuti:2005ja}
\bibitem{Benvenuti:2005ja}
  S.~Benvenuti and M.~Kruczenski,
  ``From Sasaki-Einstein spaces to quivers via BPS geodesics: $L^{p,q|r}$,''
  arXiv:hep-th/0505206.
  %%CITATION = HEP-TH 0505206;%%

%\cite{Franco:2005sm}
\bibitem{Franco:2005sm}
  S.~Franco, A.~Hanany, D.~Martelli, J.~Sparks, D.~Vegh and B.~Wecht,
  ``Gauge theories from toric geometry and brane tilings,''
  arXiv:hep-th/0505211.
  %%CITATION = HEP-TH 0505211;%%

%\cite{Butti:2005sw}
\bibitem{Butti:2005sw}
  A.~Butti, D.~Forcella and A.~Zaffaroni,
  ``The dual superconformal theory for $L^{p,q,r}$ manifolds,''
  JHEP {\bf 0509}, 018 (2005)
  [arXiv:hep-th/0505220].
  %%CITATION = HEP-TH 0505220;%%

\bibitem{kw1}
%\cite{Klebanov:1999tb}
%\bibitem{Klebanov:1999tb}
  I.~R.~Klebanov and E.~Witten,
``AdS/CFT correspondence and symmetry breaking,''
  Nucl.\ Phys.\ B {\bf 556} (1999) 89
  [arXiv:hep-th/9905104].
  %%CITATION = HEP-TH 9905104;%%

\bibitem{calabi}
E. Calabi, ``M\'etriques Kaehl\'eriennes et fibr\'es
holomorphes,'' Ann. Scient. Ec. Norm. Sup., {\bf 12} (1979), 269-294.

\bibitem{pp}
%\cite{Page:1985bq}
%\bibitem{Page:1985bq}
  D.~N.~Page and C.~N.~Pope,
``Inhomogeneous Einstein Metrics On Complex Line Bundles,''
  Class.\ Quant.\ Grav.\  {\bf 4} (1987) 213.
  %%CITATION = CQGRD,4,213;%%

\bibitem{pt}
%\cite{PandoZayas:2001iw}
%\bibitem{PandoZayas:2001iw}
  L.~A.~Pando Zayas and A.~A.~Tseytlin,
 ``3-branes on spaces with $\IR \times S^2 \times S^3$ topology,''
  Phys.\ Rev.\ D {\bf 63} (2001) 086006
  [arXiv:hep-th/0101043].
  %%CITATION = HEP-TH 0101043;%%



\bibitem{pal}
%\cite{Pal:2005mr}
%\bibitem{Pal:2005mr}
  S.~S.~Pal,
 ``A new Ricci flat geometry,''
  Phys.\ Lett.\ B {\bf 614} (2005) 201
  [arXiv:hep-th/0501012].
  %%CITATION = HEP-TH 0501012;%%

\bibitem{sfetsos}
%\cite{Sfetsos:2005kd}
%\bibitem{Sfetsos:2005kd}
  K.~Sfetsos and D.~Zoakos,
 ``Supersymmetric solutions based on $Y^{p,q}$ and $L^{p,q,r}$,''
  Phys.\ Lett.\ B {\bf 625} (2005) 135
  [arXiv:hep-th/0507169].
  %%CITATION = HEP-TH 0507169;%%

%\cite{Aharony:1997bh}
\bibitem{Aharony:1997bh}
  O.~Aharony, A.~Hanany and B.~Kol,
  ``Webs of $(p,q)$ 5-branes, five dimensional field theories and grid
  diagrams,''
  JHEP {\bf 9801}, 002 (1998)
  [arXiv:hep-th/9710116].
  %%CITATION = HEP-TH 9710116;%%

%\cite{Leung:1997tw}
\bibitem{Leung:1997tw}
  N.~C.~Leung and C.~Vafa,
  ``Branes and toric geometry,''
  Adv.\ Theor.\ Math.\ Phys.\  {\bf 2}, 91 (1998)
  [arXiv:hep-th/9711013].
  %%CITATION = HEP-TH 9711013;%%

\bibitem{MS}
%\cite{Martelli:2004wu}
%\bibitem{Martelli:2004wu}
  D.~Martelli and J.~Sparks,
 ``Toric geometry, Sasaki-Einstein manifolds and a new infinite class of
 AdS/CFT duals,''
  arXiv:hep-th/0411238.
  %%CITATION = HEP-TH 0411238;%%


\bibitem{kn}
%\cite{Klebanov:1999rd}
%\bibitem{Klebanov:1999rd}
  I.~R.~Klebanov and N.~A.~Nekrasov,
 ``Gravity duals of fractional branes and logarithmic RG flow,''
  Nucl.\ Phys.\ B {\bf 574} (2000) 263
  [arXiv:hep-th/9911096].
  %%CITATION = HEP-TH 9911096;%%

\bibitem{ks}
%\cite{Klebanov:2000hb}
%\bibitem{Klebanov:2000hb}
  I.~R.~Klebanov and M.~J.~Strassler,
``Supergravity and a confining gauge theory: Duality cascades and
chiSB-resolution of naked singularities,''
  JHEP {\bf 0008} (2000) 052
  [arXiv:hep-th/0007191].
  %%CITATION = HEP-TH 0007191;%%


\bibitem{HEK}
  C.~P.~Herzog, Q.~J.~Ejaz and I.~R.~Klebanov,
  ``Cascading RG flows from new Sasaki-Einstein manifolds,''
  JHEP {\bf 0502} (2005) 009
  [arXiv:hep-th/0412193].
  %%CITATION = HEP-TH 0412193;%%

\bibitem{ben}
%\cite{Burrington:2005zd}
%\bibitem{Burrington:2005zd}
  B.~A.~Burrington, J.~T.~Liu, M.~Mahato and L.~A.~Pando Zayas,
 ``Towards supergravity duals of chiral symmetry breaking in Sasaki-Einstein
 cascading quiver theories,''
  JHEP {\bf 0507} (2005) 019
  [arXiv:hep-th/0504155].
  %%CITATION = HEP-TH 0504155;%%

%\cite{Gepner:2005zt}
\bibitem{Gepner:2005zt}
  D.~Gepner and S.~S.~Pal,
 ``Branes in $L^{p,q,r}$,''
  Phys.\ Lett.\ B {\bf 622} (2005) 136
  [arXiv:hep-th/0505039].
  %%CITATION = HEP-TH 0505039;%%





\bibitem{kt}
%\cite{Klebanov:2000nc}
%\bibitem{Klebanov:2000nc}
  I.~R.~Klebanov and A.~A.~Tseytlin,
``Gravity duals of supersymmetric $SU(N) \times SU(N+M)$ gauge theories,''
  Nucl.\ Phys.\ B {\bf 578} (2000) 123
  [arXiv:hep-th/0002159].
  %%CITATION = HEP-TH 0002159;%%


%\cite{Dymarsky:2005xt}
\bibitem{Dymarsky:2005xt}
  A.~Dymarsky, I.~R.~Klebanov and N.~Seiberg,
  %``On the moduli space of the cascading SU(M+p) x SU(p) gauge theory,''
  arXiv:hep-th/0511254.
  %%CITATION = HEP-TH 0511254;%%

%\cite{Gubser:2004qj}
\bibitem{Gubser:2004qj}
  S.~S.~Gubser, C.~P.~Herzog and I.~R.~Klebanov,
  %``Symmetry breaking and axionic strings in the warped deformed conifold,''
  JHEP {\bf 0409}, 036 (2004)
  [arXiv:hep-th/0405282].
  %%CITATION = HEP-TH 0405282;%%



\bibitem{papa}
%\cite{Papadopoulos:2000gj}
%\bibitem{Papadopoulos:2000gj}
  G.~Papadopoulos and A.~A.~Tseytlin,
``Complex geometry of conifolds and 5-brane wrapped on 2-sphere,''
  Class.\ Quant.\ Grav.\  {\bf 18} (2001) 1333
  [arXiv:hep-th/0012034].
  %%CITATION = HEP-TH 0012034;%%

\bibitem{baryonic}
%\cite{Butti:2004pk}
%\bibitem{Butti:2004pk}
  A.~Butti, M.~Gra\~na, R.~Minasian, M.~Petrini and A.~Zaffaroni,
 ``The baryonic branch of Klebanov-Strassler solution: A supersymmetric family
 of $SU(3)$ structure backgrounds,''
  JHEP {\bf 0503} (2005) 069
  [arXiv:hep-th/0412187].
  %%CITATION = HEP-TH 0412187;%%


\bibitem{gubser}
%\cite{Gubser:1998vd}
%\bibitem{Gubser:1998vd}
  S.~S.~Gubser,
``Einstein manifolds and conformal field theories,''
  Phys.\ Rev.\ D {\bf 59} (1999) 025006
  [arXiv:hep-th/9807164].
  %%CITATION = HEP-TH 9807164;%%



\bibitem{t11spec}
%\cite{Ceresole:1999zs}
%\bibitem{Ceresole:1999zs}
  A.~Ceresole, G.~Dall'Agata, R.~D'Auria and S.~Ferrara,
 ``Spectrum of type IIB supergravity on $\mathrm{AdS}^5 \times
  T^{1,1}$ : Predictions on $\mathcal{N}  = 1$   SCFT's,''
  Phys.\ Rev.\ D {\bf 61} (2000) 066001
  [arXiv:hep-th/9905226].
  %%CITATION = HEP-TH 9905226;%%

\bibitem{resolved}
%\cite{PandoZayas:2000sq}
%\bibitem{PandoZayas:2000sq}
  L.~A.~Pando Zayas and A.~A.~Tseytlin,
 ``3-branes on resolved conifold,''
  JHEP {\bf 0011} (2000) 028
  [arXiv:hep-th/0010088].
  %%CITATION = HEP-TH 0010088;%%





%\cite{Benvenuti:2005wi}
\bibitem{Benvenuti:2005wi}
  S.~Benvenuti and A.~Hanany,
  ``Conformal manifolds for the conifold and other toric field theories,''
  JHEP {\bf 0508}, 024 (2005)
  [arXiv:hep-th/0502043].
  %%CITATION = HEP-TH 0502043;%%



\bibitem{acr}
%\cite{Arean:2004mm}
%\bibitem{Arean:2004mm}
  D.~Arean, D.~E.~Crooks and A.~V.~Ramallo,
``Supersymmetric probes on the conifold,''
  JHEP {\bf 0411} (2004) 035
  [arXiv:hep-th/0408210].
  %%CITATION = HEP-TH 0408210;%%



\bibitem{fulton}
W. Fulton, ``Introduction to Toric Varieties,'' Princeton University Press, 1993.


\bibitem{cm}
%\cite{Constable:1999ch}
%\bibitem{Constable:1999ch}
  N.~R.~Constable and R.~C.~Myers,
 ``Exotic scalar states in the AdS/CFT correspondence,''
  JHEP {\bf 9911} (1999) 020
  [arXiv:hep-th/9905081].
  %%CITATION = HEP-TH 9905081;%%

\bibitem{evans}
%\cite{Babington:2003vm}
%\bibitem{Babington:2003vm}
  J.~Babington, J.~Erdmenger, N.~J.~Evans, Z.~Guralnik and I.~Kirsch,
 ``Chiral symmetry breaking and pions in non-supersymmetric gauge /  gravity
 duals,''
  Phys.\ Rev.\ D {\bf 69} (2004) 066007
  [arXiv:hep-th/0306018].\\
  %%CITATION = HEP-TH 0306018;%%
%\cite{Evans:2004ia}
%\bibitem{Evans:2004ia}
  N.~J.~Evans and J.~P.~Shock,
 ``Chiral dynamics from AdS space,''
  Phys.\ Rev.\ D {\bf 70} (2004) 046002
  [arXiv:hep-th/0403279].\\
  %%CITATION = HEP-TH 0403279;%%
%\cite{Evans:2005ti}
%\bibitem{Evans:2005ti}
  N.~Evans, J.~Shock and T.~Waterson,
 ``D7 brane embeddings and chiral symmetry breaking,''
  JHEP {\bf 0503} (2005) 005
  [arXiv:hep-th/0502091].
  %%CITATION = HEP-TH 0502091;%%




\end{thebibliography}
\end{document}